\documentclass[a4paper, notoc]{JHEP3}
\usepackage[dvips]{graphicx}
\usepackage{amsfonts}
\usepackage{amssymb}
\usepackage{epsfig}
\usepackage{cite}

\newcommand{\be}{\begin{equation}}
\newcommand{\ee}{\end{equation}}
\newcommand{\bea}{\begin{eqnarray}}
\newcommand{\eea}{\end{eqnarray}}
\newcommand{\mbb}{\mathbb}
\newcommand{\ti}{\times}

\newcommand{\mc}{\mathcal}
\newcommand{\gsim}{\gtrsim}

\newcommand{\beqa}{\begin{eqnarray}}
\newcommand{\eeqa}{\end{eqnarray}}

 \newlength{\wth}
 \newcommand{\twographs}[2]{%
 \unitlength=1.1in
 \begin{picture}(5.8,2)
 \put(0,0){\epsfig{file=#1.eps, width=0.7\wth}}
 \put(2.5,0){\epsfig{file=#2.eps, width=0.7\wth}}
 \put(0,1.7){(a)}
 \put(2.5,1.7){(b)}
 \end{picture}
}

\title{Towards Realistic String Vacua From Branes At Singularities}
\author{Joseph P. Conlon$^{1}$, Anshuman Maharana$^{2}$,  Fernando
  Quevedo$^{2,3}$
 \\$^{1}$Rudolf Peierls Center for Theoretical Physics, 1 Keble Road, Oxford OX1 3NP, UK\\
 \\$^{2}$DAMTP,
  Centre for Mathematical
Sciences,\\
  Wilberforce Road, Cambridge, CB3 0WA, United Kingdom\\ \\
$^3$ CERN Theory Division, CH-1211, Geneva 23, Switzerland}
\abstract{
We report on progress towards constructing string models incorporating
both realistic D-brane matter content and  moduli stabilisation with
dynamical low-scale supersymmetry breaking.
The general framework is that of local D-brane models embedded into the LARGE volume
approach to moduli stabilisation. We review quiver theories on del Pezzo $n$ ($dP_n$)
singularities including both D3 and D7 branes. We provide supersymmetric examples with three
quark/lepton families and the gauge symmetries of the Standard,
Left-Right Symmetric, Pati-Salam and
Trinification models, without unwanted chiral exotics. We describe how the singularity structure leads to family symmetries
governing the Yukawa couplings which may give mass hierarchies among the different generations.
We outline how these models can be embedded into compact Calabi-Yau
compactifications with LARGE volume moduli stabilisation, and state the minimal
conditions for this to be possible.
We study the general structure
of soft supersymmetry breaking.
At the singularity all leading order contributions to the soft terms (both gravity- and anomaly-mediation) vanish.
We enumerate subleading contributions and estimate their magnitude.
We also describe
model-independent physical implications of this scenario. These include the
masses of anomalous and non-anomalous $U(1)$'s and the generic existence
of a new hyperweak force under which leptons and/or quarks could be charged. We propose that such a gauge boson could be responsible for the ghost 
muon anomaly recently found at the Tevatron's CDF detector.
}

\preprint{OUTP-08/18P\\ DAMTP-2008-101\\ CERN-PH-TH/2008-214}

\begin{document}

\tableofcontents

\section{Introduction}

String vacua aiming to describe the real world must cross various hurdles.
Among these \emph{pontes asinorum} are the requirements that the low energy particle content
incorporate the Standard Model and that the compactification geometry is stabilised with all geometric moduli being massive.
The vacuum must also break supersymmetry in such a way that Bose-Fermi splitting is not
much smaller than 1 \hbox{TeV}. While actual TeV-scale supersymmetry is not essential for
viability, it is a phenomenologically attractive feature and for our purposes we shall assume its correctness.

String theory has seen much separate effort on constructing either chiral models of particle physics or stabilised vacua.
The construction of models with a chiral matter content resembling the Standard Model dates from the earliest work
on heterotic compactifications. While no one model is compelling, the heterotic string remains a
promising arena for model-building with a steady development in the technical tools available.
Examples of recent work in this direction include \cite{hepth0512149, hepth0512177, hepth0702210,raby}.
More recently the discovery of D-branes provided new possibilities for model-building
 through intersecting branes in both IIA and IIB string theory. D-brane model building is now an extensive
 subject and is well covered by review articles such as \cite{hepth0502005, hepth0610327, fernando}.

Branes at singularities of a Calabi-Yau manifold provide an interesting class of chiral quasi realistic models.
They are local models and therefore many of their properties do not depend on the  global structure of the
compactification and are expected to survive a full compactification including moduli stabilisation. Local model-building was initiated by Aldazabal, 
Ib\'a\~nez, Quevedo and Uranga in \cite{hepth0005067}.
These authors studied models of D3 and D7 branes at orbifold singularities
including, in detail, the $\mbb{C}^3/\mbb{Z}_3 \equiv dP_0$ singularity, with the gauge group
supported on fractional D3 branes. More recent examples of local
model-building include \cite{alday,hepth0508089, 08023391, 08060634, 08053361}. In
recent years, partly motivated by the AdS/CFT correspondence,
 substantial progress has been made on the understanding
and classification of Calabi-Yau singularities, mostly on toric
singularities. General classes have been classified, such as the
$Y_{p,q}$ and $L_{a,b,c}$ singularities \cite{reviews}. Powerful techniques using
quiver and dimer diagrams have been developed that allows to go
beyond the simple orbifold singularities studied in
\cite{hepth0005067} in computing the spectrum of matter fields and the
effective superpotential. It is then worth exploring the potential
phenomenological implications for local D-brane model building of
more  general singularities.

The construction of stabilised vacua has also received much attention in recent years, and progress is reviewed in
\cite{hepth0610102, 08031194}.
Such constructions tend to require the use of fluxes and non-perturbative effects to stabilise moduli.
Arguably the best understood models are those of IIB flux compactifications, where the
fluxes stabilise the dilaton and complex
structure moduli \cite{hepth0105097} and non-perturbative effects are
required to stabilise the K\"ahler moduli.
The simplest constructions of stabilised vacua (for example KKLT \cite{hepth0301240}) are however often
supersymmetric, at relatively small volumes and with a flux
superpotential tuned to many orders of magnitude to obtain a reliable
minimum and a small
gravitino mass.
 This renders control over the $\alpha'$ expansion marginal and
makes them relatively less attractive starting points for low-energy supersymmetric
phenomenology.
These problems can
 be evaded by the LARGE volume models of
 \cite{hepth0502058, hepth0505076}. These
incorporate $\alpha'$ corrections into the K\"ahler potential and thereby generate a stable minimum
at exponentially large values of the volume.
Such models stabilise moduli deep in the geometric regime
while also generating dynamical low-scale supersymmetry breaking.

As the soft terms are induced by the moduli F-terms this falls under
the heading of gravity  mediation (or more precisely moduli mediation).
Gravity mediation occurs naturally in string theory and is attractive as
a supersymmetry breaking mechanism for its directness and its calculability.\footnote{There is a challenge of flavour non-universality
unless - as holds for example for K\"ahler moduli in
 IIB string compactifications \cite{hepth0610129,mirror} - the moduli fields responsible for supersymmetry breaking do not
 appear in the Yukawa couplings.}
In principle supersymmetry could also be communicated to the visible sector by gauge interactions. This is an interesting alternative that
 naturally gives flavour universality of soft breaking terms.
However in addition to the usual phenomenological and calculational problems (excessive CP violation,
problems with the Higgs potential and the computational difficulties of strongly coupled gauge theories), gauge
mediation in string theory is hard to realise in a controlled fashion incorporating moduli stabilisation.
 For a recent careful analysis of the potential
of realising gauge mediation in string theory, see \cite{08095064}.

An important task is to combine moduli stabilisation with realistic chiral matter sectors (for previous studies in this direction see
\cite{compact, shiu, kumar, 07110396} ). 
 Such a combination will allow a test of the assumptions that have gone into each side of this construction and may also
suggest new phenomenological possibilities.
Ideally one would hope to simply bolt together a scenario of moduli stabilisation with a D-brane MSSM-like model.
However, Blumenhagen, Moster and Plauschinn have recently in \cite{07113389} pointed out an important obstruction to this,
in that the requirement of chirality constrains the techniques used for moduli stabilisation.
Specifically, in D-brane models the chiral nature of the Standard Model implies that instantons cannot be used to stabilise the
Standard Model cycle.

The basic aims of this paper are  to make progress towards
 models which combine realistic matter sectors, full moduli
stabilisation and controlled dynamical low-energy supersymmetry breaking. The structure of this paper will be as follows.
In section \ref{sec2} we will discuss local models, as first introduced
 in \cite{hepth0005067}.
We review the philosophy of local model building and give various new
models of D3/D7 branes at del Pezzo singularities, including non-vanishing hierarchical Yukawa couplings.
In section \ref{sec3} we outline how such models can
be embedded into global moduli-stabilised compactifications, explaining how such local models may
allow the problems of \cite{07113389} to be evaded. We provide conditions on the Calabi-Yau geometry for such global embeddings to
be realised. In \ref{sec4} we discuss soft terms in this framework. Embedded into the large volume
framework, the use of branes at singularities
leads to a remarkable cancellation of all leading-order soft terms (both gravity- and anomaly-mediated).
We enumerate the possible sub-leading contributions but do not attempt a full
phenomenological analysis.

\section{Local Model Building}
\label{sec2}

\subsection{Generalities}

Phenomenological string models can be either global or local.
The basic distinction is that for local models there is a limit in which the
Standard Model gauge couplings remain finite while the bulk volume is taken to
infinity. For global models, the canonical examples of which are Calabi-Yau compactifications of the weakly coupled heterotic string,
all gauge couplings vanish in the limit that the bulk volume is taken to infinity.
We will focus on IIB string theory with D3/D7 branes, in which case the MSSM gauge interactions are supported on 7-branes
wrapping 4-cycles. In this case for local models the 4-cycles on which the MSSM is supported are vanishing cycles,
which can be collapsed to give a Calabi-Yau singularity. Local models may equally well be constructed either
at the singular locus or on the
resolution.
The simplest case has only a single resolving 4-cycle, in which case the 4-cycle is necessarily of del Pezzo type and the
singularity is a del Pezzo singularity.\footnote{We recall that the del Pezzo surfaces, $dP_n$ for $n = 0,1,2 \ldots 8$ correspond to
the blow-up into $\mbb{P}^1$s of $n$ points on $\mbb{P}^2 \equiv dP_0$.}

The use of local models is in fact forced upon us in the LARGE volume moduli stabilisation scenario of \cite{hepth0502058, hepth0505076}:
if the volume is exponentially large, the known sizes of the Standard Model gauge couplings imply that any construction of the Standard Model
is necessarily local.

Local models have various technical advantages. In global models, the chiral matter spectrum depends on the full geometry of the
compact space and cannot be computed until all global tadpoles and anomalies have been canceled. In local models, the chiral matter
is determined by only a small region of the geometry. While for consistency bulk tadpoles must still be canceled, the details of this do
not affect the chiral matter content and interactions of the local model. Local models also allow realistic matter content and coupling with a bulk 
volume
deep in the geometric regime. For global models, it has long been known that
the observed size of the Standard Model couplings implies that either the $\alpha'$ or $g_s$ expansion
is not well controlled \cite{DineSeiberg}.

One of the principal attractions of local models is their separation
between local and bulk degrees of freedom. However it is important to distinguish
between phenomenological questions that can be addressed locally and those that require some knowledge of the bulk physics.

\subsection*{What can be studied locally}

Many phenomenological quantities can be determined purely locally. These include:

\begin{enumerate}

\item The gauge groups and matter content: these are determined solely by the number of branes wrapping the local cycles and
their topological flux and intersection numbers.

\item The Wilsonian gauge couplings defined at the string scale. For D7 branes wrapping collapsible cycles, these are determined
purely by the values of the dilaton $S$, the size of the collapsing 4-cycle $T$, and the 2-forms $\int B_2$ on 2-cycles inside the collapsing 
4-cycle.
All these quantities are local.

\item The high-scale interactions between the massless modes, including Yukawa couplings.
To leading order, these are determined entirely by the local geometry and the local singularity.

\item The approximate global flavour symmetry groups, which follow purely from the local geometry.
As an example, for branes at the $\mbb{C}^3 / \mbb{Z}_3$
singularity the interactions of $(33)$ states are governed by an approximate $SU(3)$ global flavour symmetry.

\end{enumerate}

\subsection*{What can not be studied locally}

There are also many features that cannot be computed
locally and require some knowledge of the whole compactification.
In general this category includes all dimensionful scales.
The essential reason is that in string theory all
dimensionful scales derive from the string scale, which is in turn
derived from the bulk volume using
$m_s = g_s M_P/\sqrt{\mc{V}}$, and therefore requires knowledge of the global
geometry.\footnote{We work with the conventional Einstein-Hilbert general relativity action, in which
case the 4d Planck scale is fixed and the string scale is a derived quantity.} Phenomenological features
that cannot be computed in a purely local framework include:

\begin{enumerate}

\item The scale of the cosmological constant: all sectors contribute to the vacuum energy and all contributions are
additive. The answer is dominated by the size of the largest contribution.
\label{one}

\item Moduli stabilisation. Addressing the moduli problem of string compactifications
clearly requires a global approach, especially for
the closed string moduli that probe the full compactification geometry.

\item The scale of supersymmetry breaking. As for item \ref{one} above, any sector of the compactification can
break supersymmetry and contribute to supersymmetry breaking. The contributions to visible soft masses are additive and dominated
by the largest contribution.
Ensuring low-scale supersymmetry breaking requires global rather than local control.

\item The high-scale entering the phenomenological RGEs. The structure of MSSM soft terms and gauge couplings depends
crucially on the high scale from which the running starts. This is the string/Kaluza-Klein scale
and so depends on the global embedding.

\item The value of the axion decay constant. In string theory the axion-matter coupling is a
non-renormalisable coupling. The axion decay constant is typically the string scale and in any case is always
compactification-dependent.
\label{four}

\item The suppression scale for non-renormalisable operators (this includes item \ref{four} above).
 An important example of such an operator is the suppression scale $\Lambda$ entering
the quartic $\frac{1}{\Lambda} H_u H_u L L$ neutrino mass term. Depending on the particular operator,
this may be the string scale, the Planck scale, or somewhere in between.

\item Early universe cosmology, such as attempts to derive inflation and reheating or other scenarios from string theory,
 necessarily requires the dynamics of moduli stabilisation and therefore cannot be approached locally.
\end{enumerate}

The most important of these examples is probably that of supersymmetry breaking.
Viable models of supersymmetry breaking require Bose-Fermi mass splittings not smaller
than $1 \hbox{TeV}$, and models with \emph{any} source of supersymmetry breaking much larger than
this fail to provide a solution to the hierarchy problem based on supersymmetry.
In supergravity the scale of Bose-Fermi splitting is set by the gravitino mass,
$m_{3/2} = e^{K/2} W$.
In string theory the gravitino mass is a dynamical function of all fields present in the
compactification, not only those contained within the local model.
\FIGURE[r]{
\includegraphics[width=8cm]{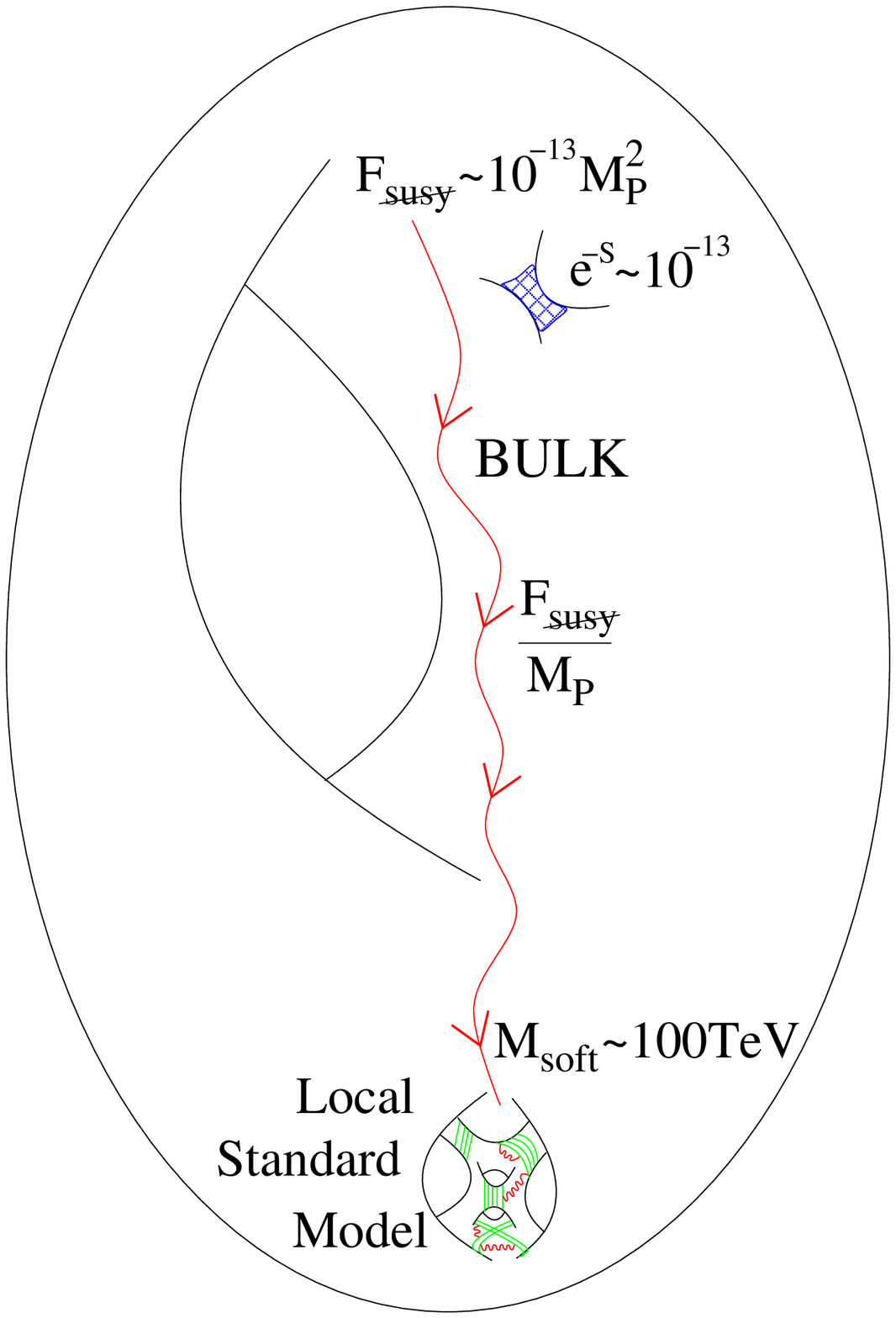}
\caption{Why it is not consistent to study supersymmetry breaking purely locally. The presence of a hidden
 D3-instanton appearing in the gauge-invariant superpotential $e^{K/2} W$ with
 amplitude $e^{-S} \sim 10^{-13}$ gives a gravitational
 contribution to Bose-Fermi splitting of order $\Delta m \sim (10^{-13} M_P) \sim 100 \hbox{TeV}$. Any such
 effect, whose presence or absence can only be determined globally,
 entirely washes out all TeV contributions of local supersymmetry breaking.
\label{globalsusybreaking}}}

For example, instantons
often generate non-perturbative contributions to the superpotential.
A single hidden-sector instanton, geometrically far separated from the local model and
with amplitude as small as $e^{-S} \sim 10^{-13} \in e^{K/2}W$, will give
a contribution to Bose-Fermi splitting one hundred times larger than that
dictated by the mass of the Higgs.
A consistent study of supersymmetry breaking therefore always requires the global compactification,
as any local model of supersymmetry breaking can be washed out by such global effects.
This point is illustrated in figure \ref{globalsusybreaking}.

We then emphasise that efforts towards a purely local description of supersymmetry breaking, such as those based on pure
gauge mediated supersymmetry breaking scenarios
can be justified only under strong assumptions
on the gravitational degrees of freedom.
For example, most gauge mediation models introduce the gravitino mass - which is a function of the moduli -
as a new \emph{ad hoc} scale $m_{3/2} \lll M_P$.
 A natural
mechanism of moduli
stabilisation at an almost Minkowski compactification without breaking supersymmetry is yet to be found.

\subsection{Branes at Singularities}

Having
restricted to the class of local models, we can further distinguish based on whether or not the local spacetime is geometric.
A supersymmetric model requires the constituent branes to be stable D-flat BPS objects that will not decay.
However the identification of such branes is well known to depend on the locus in moduli space and
may change across lines of marginal stability. In the geometric regime, the D-flatness conditions require
world-volume gauge bundles to satisfy
$J \wedge F = 0$. It is also necessary that
objects wrapping a given 4-cycle carry the same RR charge - branes and antibranes are
mutually non-supersymmetric.

In the limit of small K\"ahler moduli these conditions are modified.
At a singularity both
`branes' and `antibranes' - objects wrapped on the collapsed cycle and
carrying opposite RR charge with respect to this cycle - can be mutually supersymmetric.
In this limit the allowed interactions between supersymmetric branes is very different from the geometric limit.
In particular, for supersymmetric magnetised branes wrapped on finite-volume del Pezzos, all Yukawa couplings for the induced
chiral matter vanish \cite{08023391, 08070789}. However in the singular limit this is no longer true and
such Yukawa couplings are generically non-vanishing.

Partly for this reason, in this paper
we will focus on local model-building at the singular locus. The singular
locus also turns out to be attractive for reasons of moduli stabilisation in a global context.
We shall elaborate on this point
in sections \ref{sec3} and \ref{sec4}.

The allowed types of supersymmetric branes at a singularity
has been extensively studied.
These are the fractional branes, which come in two types.
The first type is that of fractional D3 branes (magnetised D7 branes wrapped solely on the collapsing
cycle). The second type is that of fractional D7 branes (bulk D7 branes that wrap both bulk and collapsed cycles).
All such fractional branes are wrapped on the collapsed cycle, carry twisted Ramond-Ramond charge and cannot move away from the
singularity.
They may recombine into a bulk brane, with no twisted charge, which can move away from the singularity.

The matter content for such intersecting brane models
comes in bifundamentals and is determined by the topological intersection of a pair of supersymmetric branes.
This matter content is simply expressed through a quiver
diagram. For the case of fractional D3 branes,
the quiver diagram and superpotential for $dP_0$ was computed in \cite{mrd}, that for $dP_1$, $dP_2$ and $dP_3$ in \cite{hepth0003085}
and that for $dP_4$ through $dP_8$ in \cite{hepth0212021}. The inclusion of fractional D7 branes into the quivers
is described in general, and in detail for $dP_1$, in \cite{hepth0604136}.

The matter content and superpotential of such theories may be efficiently encoded using the technology of dimer diagrams.
These also allow a simple description of the effect of introducing fractional D7 branes into the theory. We will not directly
use dimer diagrams in this paper and will instead simply write down the appropriate superpotential. The interested reader can consult
the appendix of \cite{hepth0604136} which describes dimer diagrams and in particular how they allow a general description of fractional D7 branes and 
their
interactions.

The del Pezzo spaces can be viewed as $\mbb{P}^2$ blown up at $n$ separate points.
$\mbb{P}^2$ admits an action $PGL(3, \mbb{C})$ on the projective coordinates $(z_1, z_2, z_3)$,
preserving the complex structure of $\mbb{P}^2$.
$PGL(3, \mbb{C})$ has eight complex parameters, of which two are used in fixing the position of each blow-up.
Once all parameters are exhausted the location of the blow-up represents a complex structure modulus, and thus
 $dP_n$ has $(2n - 8)$ complex structure moduli. $dP_0$ is $\mbb{P}^2$ and has the canonical
Fubini-Study metric with $SU(3)$ isometry. The isometry group is reflected in the flavour symmetry of the quiver.
As points are progressively blown up this flavour symmetry is reduced, to $SU(2) \ti U(1)$ for $dP_1$ and $U(1)$ for $dP_2$.
For higher del Pezzos, there are no flavour symmetries of the superpotential, and for $n > 4$,
the superpotential (and thus the Yukawa couplings) depend on the complex structure moduli.

In principle the MSSM may arise from
any configuration of supersymmetric branes at any singularity. However, there are many singularities
and a global search may not be most productive. We shall organise our analysis using two general principles:
triplication, and the presence of flavour symmetries.
The three Standard Model families make triplication of matter content essential.
Flavour symmetries are also desirable. One of the most
striking features of the Standard Model is the pattern of Yukawa couplings. While the origin of the Yukawas is
unknown, one attractive idea is that the Yukawas are governed by
approximate flavour symmetries under which different generations take different charges.
Flavour symmetries are also appealing in models of
neutrino masses and supersymmetry breaking.

For this reason, we shall mostly focus on the lower-degree del Pezzos, which automatically generate family
symmetries.
The $dP_0$ singularity is simply $\mbb{C}^3 / \mbb{Z}_3$, with a manifest $SU(3)$ global symmetry.
In fact we shall see that $SU(3)$ is too large as a family symmetry and gives problematic Yukawas. For this reason models based on
$dP_1$ or $dP_2$ are more attractive. $dP_1$ has an $SU(2) \ti U(1)$ family symmetry.
This symmetry is also shared by $Y^{P,Q}$ singularities.
However, unlike the del Pezzo case these do not give rise to family triplication.\footnote{We thank A. Uranga for very useful discussions on this 
subject.}

As shown in \cite{hepth0005067},
it is easy to construct models on $dP_0$ with realistic spectra, with hypercharge emerging naturally as the unique anomaly-free $U(1)$.
We start by reviewing the structure of these models, before describing how they can be generalised
to the more attractive $dP_1$ case. Some of the following models have already appeared in
 \cite{hepth0005067} and others are new.

\subsection{Del Pezzo 0}
\label{dP0sec}

\begin{figure}[h!]
\centering{\includegraphics[height=5cm]{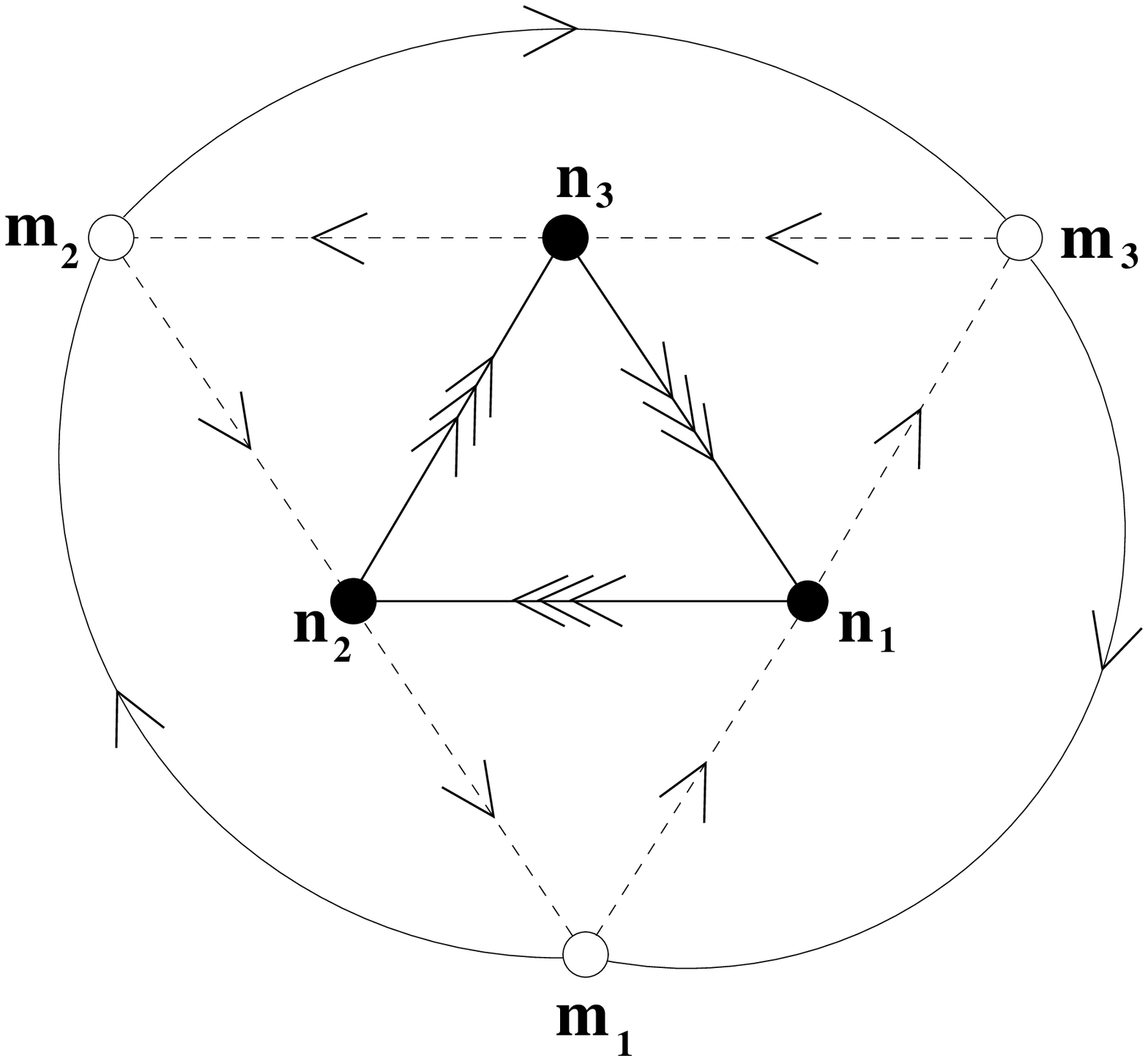}}
\caption{The quiver for the $dP_0$ singularity. Dark circles correspond to fractional D3 branes wrapping only the collapsed cycles and support
 Standard Model gauge groups.
White circles correspond to fractional D7 branes wrapping both bulk and collapsed cycles and support bulk hidden sector gauge groups.
Standard Model matter arises from either $D3-D3$ or $D3-D7$ states.}
\label{dP0}
\end{figure}

The full $dP_0 \equiv \mbb{C}^3/\mbb{Z}_3$ quiver, including the
possible presence of fractional D7 branes, is shown in figure \ref{dP0}. This quiver has been studied extensively,
using the language of Chan-Paton factors, in the paper
\cite{hepth0005067}.
An important general point is that,
as a bifundamnetal under non-Abelian gauge groups,
the $Q_L$ fields must exist as one of the internal $33$ lines in the quiver.


The $\mbb{C}^3/\mbb{Z}_3$ geometry
has a manifest $SU(3)$ symmetry under $z_i \to U_{ij} z_j$.
This global $SU(3)$ symmetry is reflected in the superpotential for the $33$ interactions, which is
\be
W = \epsilon_{ijk} X_i Y_j Z_k.
\ee
This superpotential has an $SU(3)$ flavour symmetry,
under which $X$, $Y$ and $Z$ all transform as $\bf{3}$s, with the superpotential corresponding to the baryonic $SU(3)$ invariant.
As a symmetry of the full Lagrangian the $SU(3)$ flavour symmetry is
however broken by the presence of fractional 7-branes. Each 7-brane singles out a complex plane and
thus breaks the flavour symmetry to $SU(2) \ti U(1)$.

A fractional 7-brane is defined by its Chan-Paton factor and the bulk cycle it wraps.
The Chan-Paton factor corresponds to the magnetic flux of the 7-brane on the collapsed cycles.
This determines the intersection numbers
with the fractional 3-branes and thus the matter content. In figure \ref{dP0} the different
white circles correspond to different choices of
Chan-Paton factor for the 7-brane. The choice of bulk cycle wrapped by the 7-brane does not affect the matter content but does affect the 
superpotential.
The superpotential for (33)(37)(73) interactions is
\be
W = \Phi_{33}^i \Phi_{37_i} \Phi_{7_i 3}.
\ee
That is, a $7_i$-brane (one not wrapping the $i$th complex dimension) couples only to
the $33$ state along the $i$th complex dimension. The full superpotential for
$dP_0$ is therefore
\be
W = \epsilon_{ijk} \Phi_{33}^i \Phi_{33}^j \Phi_{33}^k + \sum \Phi_{33}^i \Phi_{3 7_i} \Phi_{7_i 3},
\ee
where we have suppressed all gauge indices. Note that by the choice of the bulk cycle and Chan-Paton factor for the 7-brane,
a unique (33) state is singled out which interacts with the (37) states.
We also note that the (33) interactions respect the full $SU(3)$ symmetry,
whereas the (33)(37)(73) interactions respect only the smaller
$SU(2) \ti U(1)$ symmetry preserved by the 7-brane.
With sufficient D7-branes, the $SU(3)$ symmetry is completely broken as a symmetry of the full Lagrangian.

For generality, we first allow an
arbitrary number of branes on each node, labelled $n_i$ for the D3
branes and $m_j$ for D7 branes as in figure \ref{dP0}. The gauge
theory carried by the D3 brane nodes is $U(n_1)\times U(n_2)\times
U(n_3)$. For the D7 branes, the gauge group depends on the bulk cycles
and we leave this open.\footnote{Each white circle can be split into three separate nodes, one for
each choice of bulk 4-cycle. The D7 gauge group depends on the details of this splitting.} The $i$-th D7 node will
correspond to a subgroup of $U(m_i)$. Since the
standard model gauge group must come from the D3 brane sector,
we leave the D7 brane groups unspecified and only count
the multiplicity from the number of D7 branes on each node.

The chiral matter spectrum under $SU(n_1)\times SU(n_2)\times
SU(n_3)$ can be written as:
\bea
3\left[ \left(\bf{n_1}, \bf{\bar{n}_2}, \bf{1} \right) +
  \left(\bf{1},\bf{n_2}, \bf{\bar{n}_3}\right) + \left(\bf{\bar{n}_1},
    \bf{1}, \bf{n_3} \right) \right]  + m_1\left[\left(\bf{\bar{n}_1},
      \bf{1}, \bf{1}\right) + \left(\bf{1}, \bf{n_2}, \bf{1} \right) \right]
\nonumber \\
+m_2\left[\left(\bf{1}, \bf{\bar{n}_2}, \bf{1}\right) + \left(\bf{1},
  \bf{1}, \bf{n_3}\right)\right] + m_3\left[\left(\bf{1}, \bf{1},
  \bf{\bar{n}_3}\right) + \left(\bf{n_1}, \bf{1}, \bf{1} \right)
  \right]
\eea
The quantum numbers under the $U(1)$ factors  of $U(n_i)=SU(n_i)\times
  U(1)$ are $+1$ for a fundamental, $-1$ for an
  antifundamental and $0$ for a singlet.

To these particles we must add D3 brane singlets from the
intersections among different D7 branes. These are
particles which will not be charged under the standard model group and appear in
the outer circle of figure \ref{dP0}. A
non-vanishing vev breaks the D7 gauge symmetries and gives masses to
D3-D7 states. As remarked in \cite{hepth0005067},
if the standard model comes from the D3 brane sector of the quiver
diagram, the $dP_0$ models will naturally lead to three families and
at most three non-abelian factors.

The consistency requirement of
tadpole cancellation implies anomaly cancellation for all non-abelian
gauge symmetries. This equates
the number of fundamentals and antifundamentals
for all nodes on the quiver.\footnote{In quiver diagrams $SU(2)$ nodes are `really' $U(2)$.
By a slight abuse of notation, we use $\bf{2}$ and $\bar{\bf{2}}$ to refer to $SU(2)$ $\bf{2}$s with opposite
charges under the $U(1)$ of $U(2) = SU(2) \ti U(1)$. Likewise we refer to the $\bf{2}$ and $\bar{\bf{2}}$ as fundamentals and
antifundamentals of $SU(2)$.} The number of D7 branes is therefore given by:
\be
m_2= 3\left(n_3-n_1\right) +m_1 \qquad m_3= 3\left(n_3-n_2\right) + m_1,
\label{cons}
\ee
with the constraint $m_{i} \geq 0 $ imposed.
The complete set of solutions is obtained  by ordering the $n_{i}$
as $n_{3} \geq n_{2} \geq n_{1}$ (without loss of generality) and taking $m_{1} \geq 0$.
This determines $m_{2,3}$ by (\ref{cons}).

There are two anomalous $U(1)$s, which are cancelled by the
Green-Schwarz terms induced by the integrals
of RR forms of the form $\int_{\gamma_4} C_4$ and $\int_{\gamma_2}
C_2$,
with $\gamma_{4,2}$ the associated $4$ and $2$ cycles.
These are local modes in the sense that they have normalisable kinetic terms in the non-compact limit.
This follows from the fact that the local 2- and 4-cycle of $dP_0$ are dual to each other (so the $dP_0$ has non-zero
self-intersection). As will be shown in section \ref{sec3},
the two anomalous $U(1)$'s both receive masses at the string scale, $m \sim \frac{1}{\sqrt{\alpha'}}$.
The unique anomaly-free $U(1)$ is
\be
Q_{anomaly-free} = -\sum_{i=1}^3 \frac{Q_i}{n_i},
\ee
where $n_i$ is the rank of the $i$th gauge group factor and $Q_i$ the diagonal $U(1)$ of this factor.

Let us now discuss some phenomenologically attractive models where
the D3 brane gauge group corresponds to the Standard Model, the
Left-Right Symmetric Model, the Trinification Model and the Pati-Salam model.

\subsubsection{Standard Model}

\begin{figure}[h!]
\centering{\includegraphics[height=5cm]{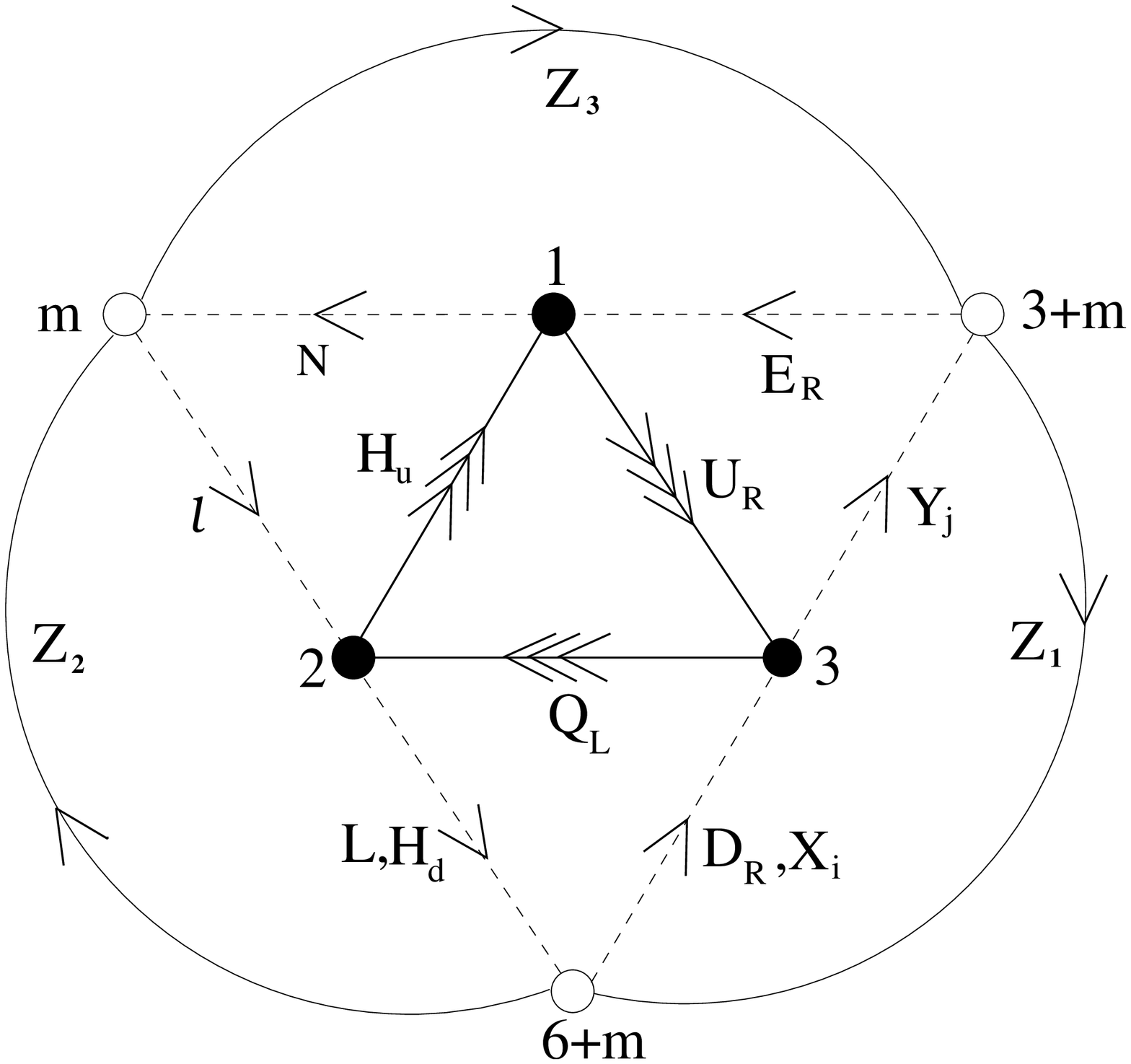}}
\caption{The quiver for the Standard Model realised at a $dP_0$ singularity.}
\label{dP0SM}
\end{figure}
This is a slightly generalised version of the models already discussed
in \cite{aiq, hepth0005067}. The spectrum can be seen in the quiver diagram
\ref{dP0SM}. Some features should be emphasised:

\begin{itemize}
\item{} The total number of D7 branes is determined by the free
   parameter $m_{1}$. The simplest case $m_1$ reproduces the models
  in \cite{hepth0005067}.

\item{} As expected, the unique non-anomalous $U(1)$, $Q_{anomaly-free}$  is
  precisely hypercharge. However, the normalisation is not standard \cite{aiq}. Using the standard normalisation for the $U(n)$
generators to be $\rm{Tr} T^2=1/2$
the hypercharge normalisation is $k_1=11/3$ different from the standard GUT normalisation $k_1=5/3$.
 This gives the Weinberg angle $\sin^2\theta_w=1/(1+k_1) =3/14=0.214$ already close to the experimental one ($\sim 0.2397$)
indicating that loop corrections, unlike the GUT case, should be small.

\item{} The left handed quarks $Q_L$ together with the right handed up
  quarks $U_R$ and the down Higgs $H_u$ come in three copies and
  couple with a superpotential $\epsilon_{ijk} Q_L^i U_R^j H_u^k$
  which gives masses to up-quarks. On giving a vev to the Higgs field the quark mass matrix is
  $$
  M_{ij} = \left( \begin{array}{ccc} 0 & M & 0 \\ -M & 0 & 0 \\ 0 & 0 & 0 \end{array} \right),
  $$
with two heavy quarks and one light one.

\item{} Both the right handed down quarks $D_R$ and the (three) down Higgs $H_d$
are D3-D7 states. The allowed coupling $Q_L D_R H_d$
  provides masses to down quarks. There are $(m+3)$
extra $SU(3)$ vector-like triplets  $X_i, Y_i$ that in principle can obtain
  a mass if the standard model singlets $Z_1$ get a non-vabishing vev.

\item{} All leptons are D3-D7 states.
 The left-handed ones leptons $L$ have
  the same origin as the down Higgsses but couple to $Q_L$ and $X$ as
$Q_L L X$.
If $X$ is heavy and integrated out of the low-energy
  effective theory this interaction is not relevant for low-energy physics.
The $3+m$ right-handed electrons $E_R$ couple to $U_R$ and
  $Y$. There are $m$ extra fields $N, l$ that couple to $H_u$.
Finally there are no clear identifiable right-handed neutrinos.
  These could come from the standard model singlets $Z_{1,2,3}$ or
  other heavy singlets such as Kaluza-Klein excitations of moduli fields.

\item{} If the standard model
singlets $Z_{1,2,3}$ get a vev the spectrum reduces to
three copies of: $Q_L, U_R, D_R, L, E_R, H_u, H_d$ which is precisely
the MSSM spectrum (with all right quantum numbers including
hypercharge)
 plus two extra Higgs pairs. Yukawa couplings are induced for both up
 and down quarks but not for leptons.

\item{}  If the blow up mode is  stabilised at the singularity all dangerous $R$-parity
 violating operators are forbidden by a combination of the global symmetries descending from
anomalous $U(1)$'s \cite{iq} (see also \cite{aiq,ak}).
\footnote{In particular, anomalous $U(1)$ at the node 3 in figure \ref{dP0SM} correponds to baryon number.
 This mechanism for proton stability seems to be a generic feature
of models constructed from quivers.} As in the Standard Model, such symmetries can be broken by
 non-perturbative effects, but these are usually suppressed.

If the Fayet-Iliopoulos (FI) parameter (blow-up mode) is stabilised at a non-zero value, in general
 R-parity violating operators will appear in the  effective action.
For small values of the blow-up mode one expects the coefficients of these operators to be suppressed by
powers of the blow-up vev  in string units. These operators would
induce proton decay through sfermion exchange with a rate
\begin{equation}
    \Gamma \sim  \bigg(   {   |\phi| \over M_{\rm string} } \bigg)^{2(p+q)}    {   m^{5}_{\rm proton} \over 16 \pi^{2} {  M^{4}_{\rm susy}} }
\end{equation}
where $M_{\rm susy}$ is the SUSY breaking scale, $\phi$ the vev of the blow up, and $p$ and $q$ the suppression
powers  of the two MSSM vertices involved in the process . Comparing to the current bounds of $10^{32}$ years
for the proton lifetime, we require
\begin{equation}
   \bigg({   \langle|\phi| \rangle\over M_{\rm string} } \bigg) ^{(p+q)} < 10^{-27}
\end{equation}
 For the LVS, taking
$M_{\rm string} = 10^{12} {\rm GeV}$, for $|\phi| \sim 10 { \rm TeV}$ the bound implies $p+q \geq 4$.

Another possibility is that the symmetry breaking process leaves some remaining discrete symmetries that forbid
$R$-parity violating operators. In \cite{aiq} a concrete example was found in which a $\mbb{Z}_2$ symmetry
 coming from  the fact that D3-D7 states couple in pairs combines with a
  remnant
$\mbb{Z}_2$ from the breaking of the gauge symmetry to give rise to an effective $R$-parity.
 \end{itemize}

\subsubsection{Left-Right Symmetric Models}

\begin{figure}[h!]
\centering{\includegraphics[height=5cm]{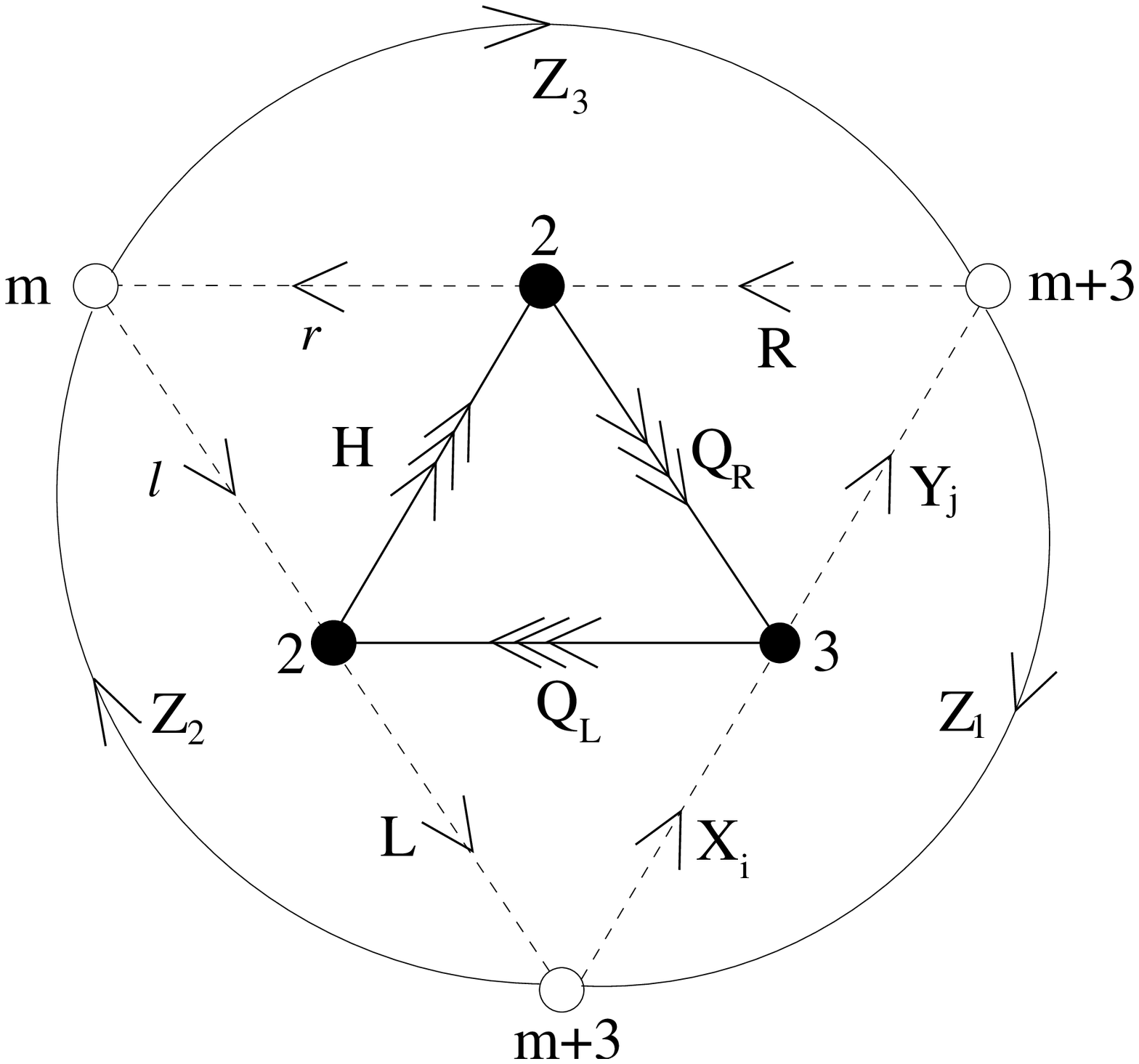}}
\caption{The quiver for the Left-Right symmetric model
realised at a $dP_0$ singularity.}
\label{dP0LR}
\end{figure}

A second simple class of models are the left-right symmetric models with
gauge symmetry $SU(3)_c\times SU(2)_L\times SU(2)_R \times U(1)_{B-L}$
 previously studied in
\cite{aiq, hepth0005067}. Figure \ref{dP0LR} represents the general class of these
models.
These models offer an interesting generalisation of the standard-like
models.

\begin{itemize}

\item{} The anomaly free combination $Q_{anomaly-free}$ is
  $U(1)_{B-L}$, with normalisation $k_{B-L}=32/3$. Upon breaking to the standard model this leaves to the same Weinberg angle as before.

\item{} The D3-D3 sector gives three families of both left-handed
  quarks $Q_L$ and right-handed quarks $U_R, D_R$ which come in an $SU(2)_R$ doublet
$Q_R$.  Both Higgsses $H_u, H_d$ contain another $SU(2)_R$
doublet $H$ and also come in three families. Unlike the Standard Model case, they are clearly distinguished from leptons.
The Yukawa couplings for all quarks come from the
coupling $\epsilon_{ijk} Q_L^i Q_R^j H^k$.

\item{}
The $(m+3)$ leptons $L,R$ are  in the D3-D7 sector with no Yukawa couplings.
The leptons $R$ include both the $E_R$ and the right-handed neutrinos $\nu_R$.

\item{} There are $m+3$ pairs of vector-like triplets $X,Y$ that can get a
  mass if the LR singlets $Z_1$ get a vev.

\item{} The  $n$ extra D3-D7 fields, $r,l$ couple to the Higgsses as
  $Hrl$. These can also be made massive by giving a vev to the singlets $Z_{2,3}$.

\item{} If all $Z_{1,2,3}$ get a vev, the model reduces to simply the
  supersymmetric version of the LR model plus two extra Higgsses.

\item{} A nonvanishing vev for the fields $R$ induces
the breaking of $SU(2)_R\times U(1)_{B-L} \rightarrow U(1)_Y$. Here
hypercharge $Y=T_R + Q_{B-L}$ and $T_R$ is the $U(1)$ generator inside
$SU(2)_R$.
This symmetry breaking should be at a similar scale as the Standard
Model symmetry breaking ($\langle R\rangle \gsim \langle H\rangle$)
and is expected to be induced after supersymmetry breaking.

\item{} $U(1)_{B-L}$ prevents the proton from decaying and the
  symmetry can survive as a global symmetry if the blow-up mode is
  stabilised at the singularity.

\item{}In references \cite{aiq, hepth0005067} it was found that this class of
  models leads to gauge coupling unification at the
  intermediate scale $\sim 10^{12} $ GeV with the same level of
  precision as the MSSM. It is interesting to notice
  that this is also the scale preferred from the
LARGE volume scenario of moduli stabilisation in order
  to have TeV scale of soft supersymmetry breaking terms.

\item{} In \cite{hepth0005067} an extension of this model to a (singular) F-theory compact model with the LR symmetric model living inside 7 D3 
branes and 6
D7 branes at a $\mathbb{Z}_3$ singularity. This was the first realistic supersymmetric {\it compact} D-brane model constructed explicitly and serves 
as
the prime example that the bottom-up approach of local model building can actually be embedded in compact Calabi-Yau constructions with all
tadpoles cancelled. (For other constructions of compact models including warped throats see \cite{compact}.)
Unfortunately, for our purposes, this compactification does not seem to satisfy the conditions for a LARGE volume compactification and F-theory at
singularities is yet to be properly understood \cite{WU}.

\end{itemize}

\subsubsection{Pati-Salam Models}

\begin{figure}[h!]
\centering{\includegraphics[height=5cm]{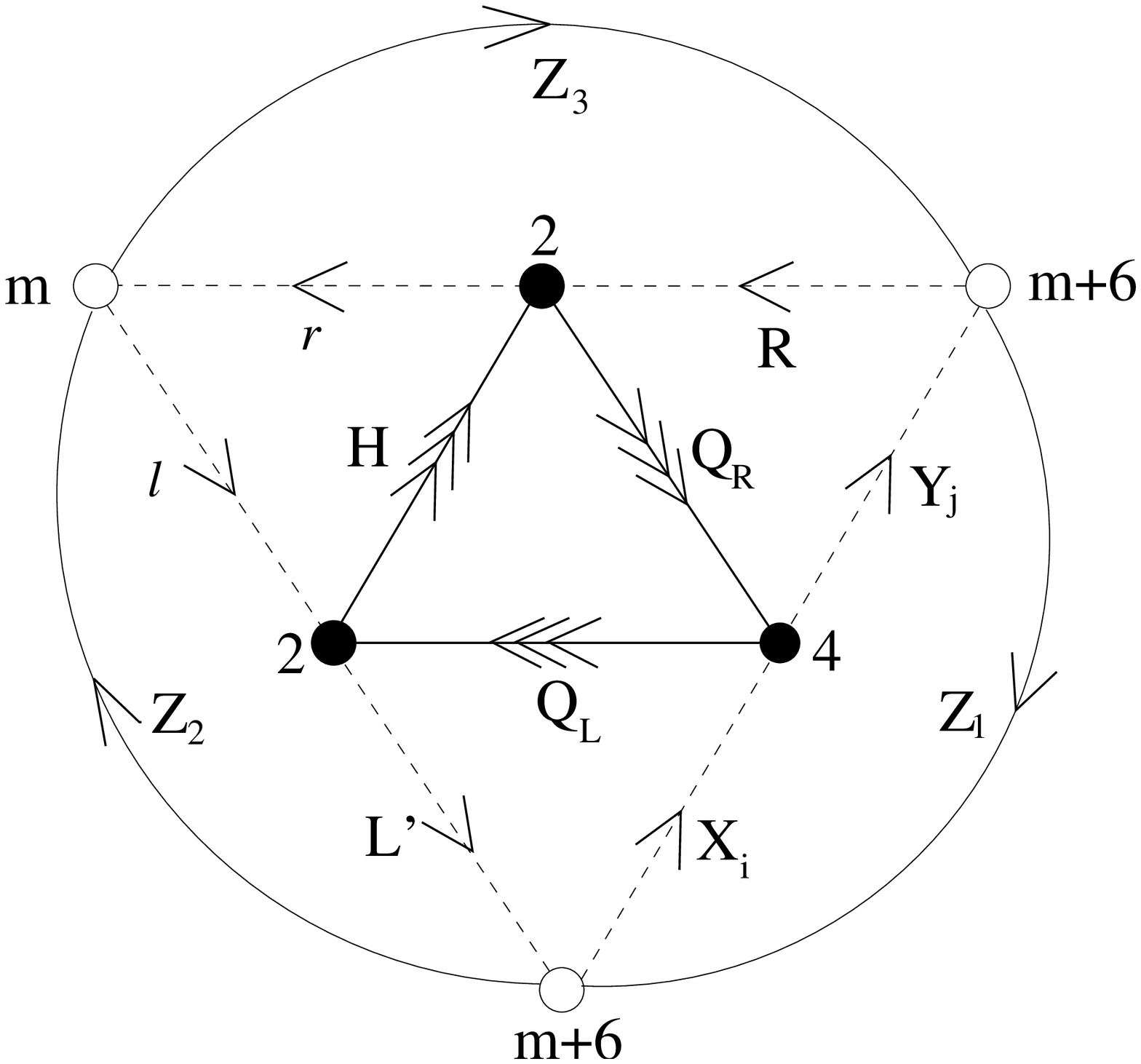}}
\caption{The quiver diagram for the Pati-Salam models
realised at a $dP_0$ singularity.}
\label{dP0PS}
\end{figure}

The natural next step is to costruct Pati-Salam models with three
families of
$SU(4)\times SU(2)_L \times SU(2)_R\times U(1)$.
These are illustrated in the quiver diagram \ref{dP0PS}.

The main ingredients of these models are.

\begin{itemize}
\item{}
All 16 standard model particles, including the right handed neutrinos,
fit precisely in the D3-D3 part of the spectrum as in a full $\bf{16}$
of $SO(10)$.
In particular the field $Q_L$ transforming in the $(\bf{4}, \bf{\bar{2}}, \bf{1})$
includes left handed quarks and leptons. This is remarkable and in principle appears as
a substantial advantage over the previous models.
Yukawa couplings for all quarks and leptons may be generated from the
superpotential
$\epsilon_{ijk} Q_L^i Q_R^j H^k$.

\item{} The scalar right-handed neutrino inside the $(\bf{\bar{4}},
  \bf{1}, \bf{2})$ may participate in the breaking of the symmetry to
  the standard model. This would however give a mass to some of the
  Higgses and leptons.

\item{}
There are extra doublets of both $SU(2)$'s ($L', l, r, R$) from the D3-D7 sector, and also
(anti) fundamentals of $SU(4)$, $X$ and $Y$. These in principle
could be used to break $SU(4)\times U(1)$ to $SU(3)_c\times
U(1)_{B-L}$. They can also become heavy if the $Z_1$ fields get a
vev. The fields $r,R$ can be used to break $SU(2)_R\times U(1)_{B-L}$
to $U(1)_Y$.

\item{} If the $Z_{1,2,3}$ fields get a vev we would be left with the
  three families of the original Pati-Salam model together with 6
  copies of the left(right) doublets $L' (R)$.

\end{itemize}

\subsubsection{Trinification Models}

Another interesting extension of the Standard Model is the
trinification model with three families of $SU(3)^3$ as shown in the figure.

\begin{figure}
\centering{\includegraphics[height=5cm]{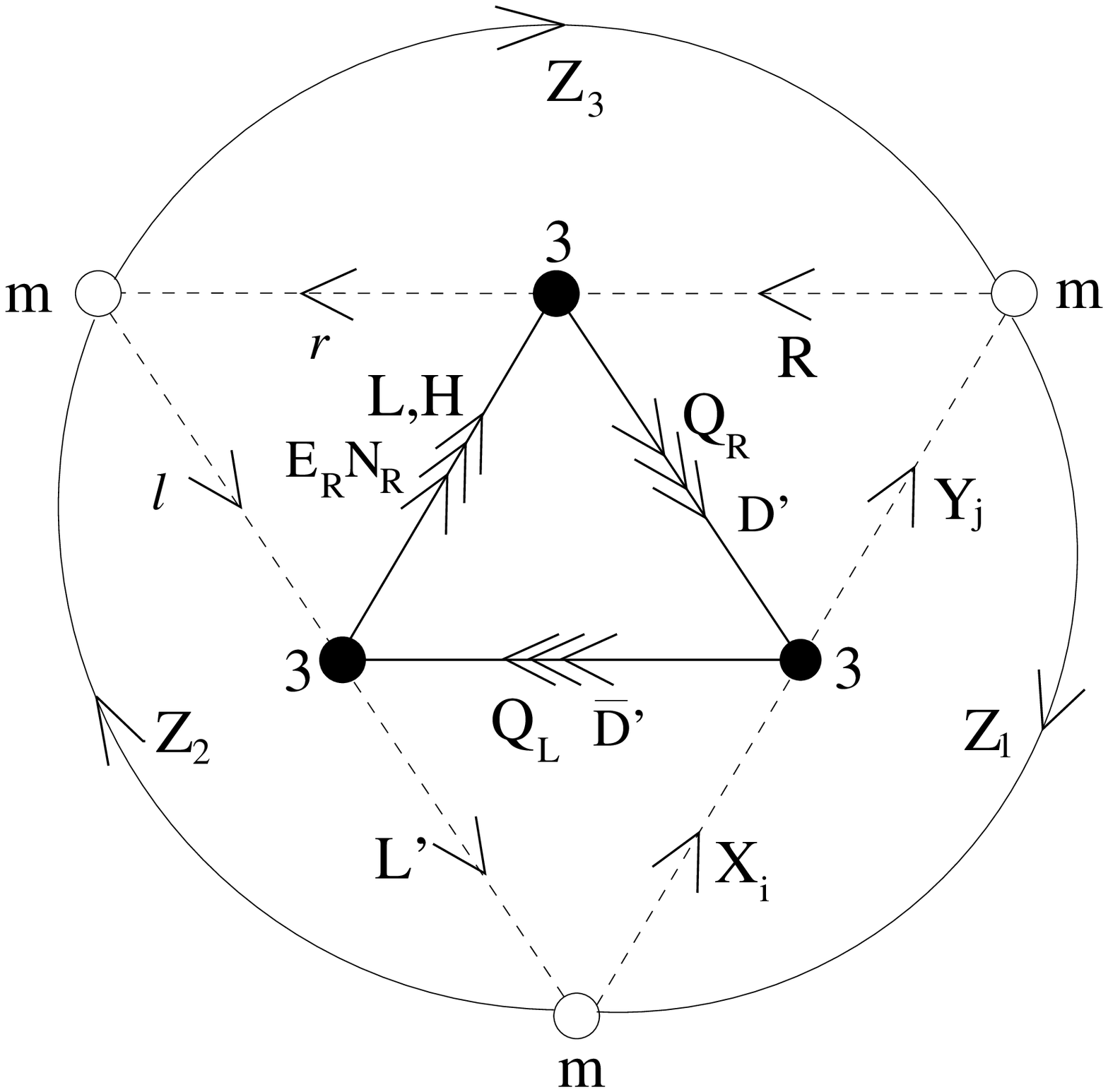}}
\caption{Quiver diagram for the Trinification Models. }
\label{delPezzoTrinModels}
\end{figure}

\begin{itemize}

\item{}
The anomaly free $U(1)$, $Q_{anomaly-free}$ is in this case a trivial
overall $U(1)$ that decouples. So in this case the model is $SU(3)_c\times
SU(3)_L\times SU(3)_R$ and there are no extra massless $U(1)$'s.
In this case the origin of hypercharge
has to be from the rank-reduction breaking of $SU(3)_L \times SU(3)_R$.

\item{}
These models are particularly simple as, since all nodes in the quiver
diagram are equal, there is no requirement to add D7 branes to
cancel the anomalies. All the standard model particles, plus additional matter,
 fit in the 27 states of the D3 brane sector:
\be
3 [ \left(\bf{3}, \bf{\bar{3}}, \bf{1}\right) + \left( \bf{1}, \bf{3},
  \bf{\bar{3}} \right) + \left( \bf{\bar{3}}, \bf{1}, \bf{3} \right) ]
\ee
which corresponds to a $\bf{27}$ of $E_6$. Therefore this model is
similar to a a Calabi-Yau compactification with three families of
$\bf{27}$'s after the breaking $E_6\rightarrow SU(3)_c\times
SU(3)_L\times SU(3)_R$ \cite{ross}.
The first nine states
include the left-handed quarks $Q_L$ plus one (exotic)
triplet $\bar{D}'$
of hypercharge $Y=-1/3$. The second nine states include the right
handed quarks plus an extra down quark, $D'$. The rest include the
leptons and Higgsess including two right-handed neutrinos. A vev
for the scalar components of the right-handed neutrinos can break the
symmetry to the standard model giving also a mass to the extra
triplets
$D', \bar{D}'$. However, in this process the would-be leptons are
Goldstone-bosons that are eaten by the gauge fields.

\item{} The D3-D7 spectrum consists of a number $m$ of
pairs of $\bf{3}$ and
  $\bf{\bar{3}}$ for each of the $SU(3)$ gauge groups and could play a
  role for gauge symmetry breaking, provided they do not all receive a mass
  by vev's of the $Z_i$ fields.
\end{itemize}

In summary, while none of these models are fully realistic, we have a series of
interesting models with three families all containing the matter content of the MSSM and no chiral exotics.
There are further models that can easily be considered, for example the
$331$ model for which only one sector of D7's is needed for anomaly
cancellation. Furthermore, as in \cite{hepth0005067}, we could have considered models at orientifold singularities obtaining
for instance a three-family $SU(5)$ model and its extensions to higher del Pezzo singularities.
A detailed analysis of the phenomenological
prospects of each model is out of the scope of this article.

One general problem we note is that
anomalous $U(1)$s tend to forbid the existence of
Yukawa couplings for leptons. This is because
the leptons $L$ and $E_R$
come from different D3-D7 sectors, and the orientation of the arrows
(which indicate the $U(1)$ charge of $U(n)=SU(n)\times U(1)$) do
not allow a non-vanishing coupling among them. This is less of an issue for the Pati-Salam and Trinification models,
where Standard Model fields are all D3-D3 states. However in this case it is difficult to break the gauge groups down to the
Standard Model.

We shall however concentrate on one general issue regarding
Yukawa couplings that applies to all the models based around $dP_0$.
While the matter content of the models is appealing, the $SU(3)$ global symmetry of
the $33$ sector is always problematic.
Once one of the Higgs fields acquires a vev, the Yukawa matrix can be written without loss of generality as
$$
Y_{ijk} \sim \left( \begin{array}{ccc} 0 & M & 0 \\ -M & 0 & 0 \\0 & 0 & 0 \end{array} \right).
$$
This mass matrix can be diagonalised as $(M, M, 0)$, and therefore all models based around
figure \ref{dP0} make the unacceptable prediction that
$m_t \sim m_c$.

\subsection{Del Pezzo 1}
\label{dP1sec}

The origin of the problematic Yukawa texture for models based on $dP_0$ was
the over-large global $SU(3)$ family symmetry.
We want to keep the many attractive features of the
$dP_0$ models while reducing this family symmetry.
As the size of the symmetry group is reduced with the height of the del Pezzo,
this naturally leads us to higher del Pezzos. However, as $n$ increases the family symmetry of the quiver disappears entirely.
As flavour symmetries are phenomenologically attractive and we prefer to maintain them, we therefore focus on models based on
$dP_1$. The quivers for lower degree del Pezzos can be obtained from higgsing higher del Pezzos and
so the models we now describe can be naturally generalied to $dP_{n>1}$.

\begin{figure}[h!]
\centering{\includegraphics[height=5cm]{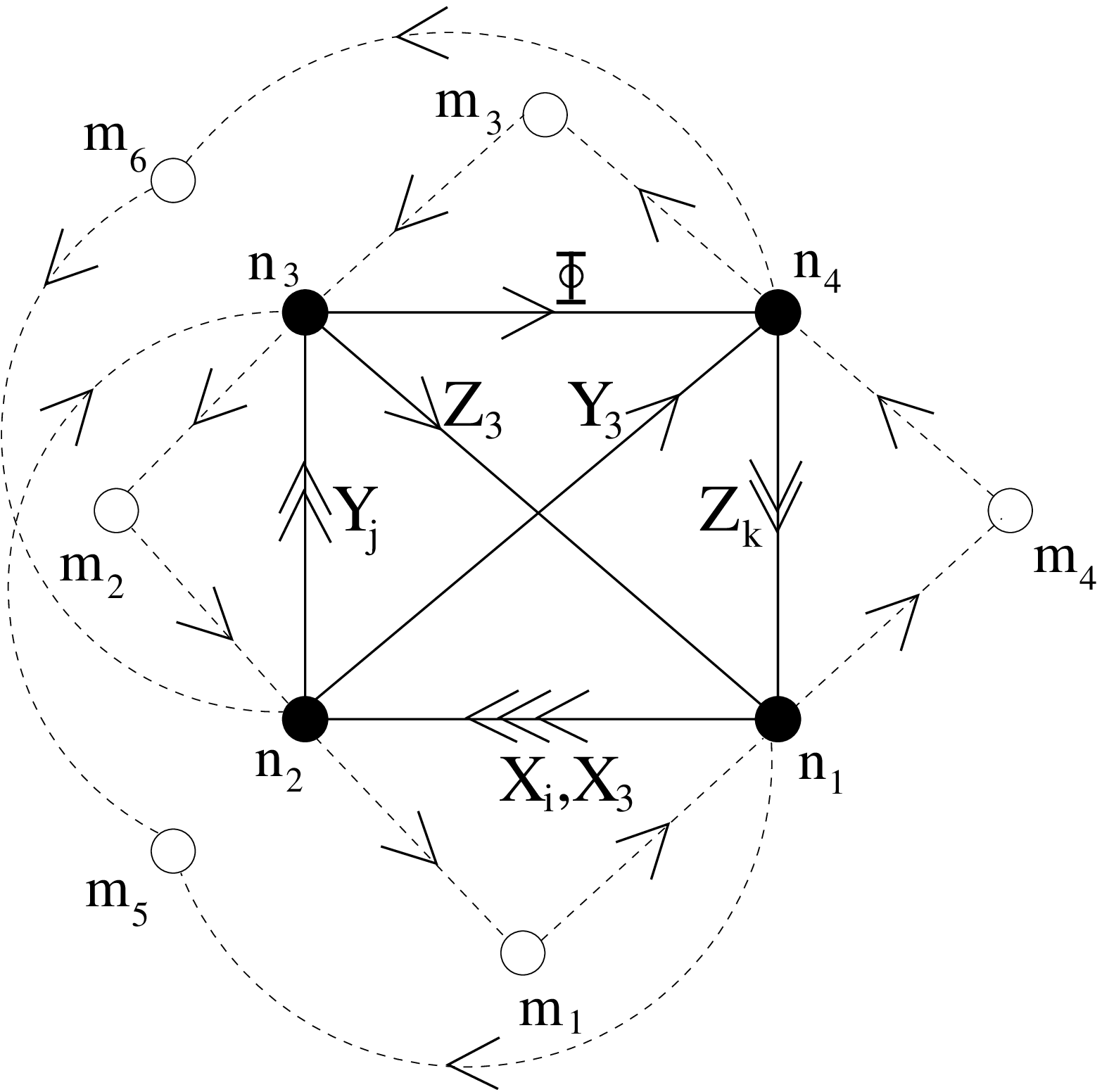}}
\caption{The $dP_1$ quiver, including all possible fractional D7 branes. Black circles denote fractional D3 branes and
white circles fractional D7 branes. We have only shown $33$ and $37$ states.}
\label{delPezzo1}
\end{figure}

The $dP_1$ singularity is not an orbifold but is toric,
and can be obtained through successive blow-ups of the $\mbb{C}^3/\mbb{Z}_3 \ti \mbb{Z}_3$ orbifold singularity \cite{hepth0003085}.
The allowed spectrum of fractional D7 branes for this model were computed in \cite{hepth0604136}.
The quiver for this theory, including the possible supersymmetric fractional
D7 brane states, are shown in figure \ref{delPezzo1}.
For every $33$ state $\Phi_{3_i 3_j}$, there exists a supersymmetric 7-brane giving an $(7i)$ fundamental and a
$(7j)$ antifundamental with the Yukawa coupling $\Phi_{3_i 3_j} (7i) (7j)$.

The superpotential for the 33 states for the $dP_1$ quiver is
\be
W = \epsilon_{ij}X_i Y_j Z_3 - \epsilon_{ij} X_i Y_3 Z_j + \frac{\Phi}{\Lambda} X_3 \epsilon_{ij} Y_i Z_j,
\label{sp}
\ee
where $\Lambda$ is an appropriate UV cutoff.\footnote{Note that within the low energy $\mc{N}=1$ supergravity,
$\Lambda$ is necessarily $M_P$ due to holomorphy, and the actual physical suppression scale is determined by terms in the K\"ahler potential.}
There is an $SU(2)$ flavour symmetry under which $X$, $Y$ and $Z$ transform as $\bf{2}$s, and also a $U(1)$ flavour symmetry under which
$X_3$ has charge $+1$ and $\Phi$ charge $-1$. There is also a $U(1)_R$ symmetry and the three $U(1)$s descending from the four D3 brane vertices, 
with
an overall $U(1)$ decoupling.

\vspace{0.5cm}

\begin{table}
\begin{center}
\begin{tabular}{ | c| c|}
\hline
 singularity  & flavour symmetry  \\
\hline
$dP_{0}$ & SU(3)  \\
\hline
$dP_{1}$ & SU(2) $\times$ U(1) \\
\hline
$dP_{2}$ & $U(1)$ \\
\hline
$dP_{n>2}$ & \rm{none} \\
\hline
\end{tabular}
\caption{Table showing the continuous flavour symmetry associated with 33 states for various del Pezzo singularities.}
\end{center}
\end{table}

As with the $dP_0$ quiver, fractional 7-branes can be inserted which wrap both bulk and collapsed cycles.
Analogously to $dP_0$, and as described in detail in the appendix of \cite{hepth0604136}, there is one 7-brane for each (33) state.
For every (33) state, there is a 7-brane that leads to a (33)(37)(73) Yukawa coupling coupling only to that (33) state. At the level of
gauge interactions, this is visible in the presence of the white dots in figure \ref{delPezzo1}, which lead to Yukawa couplings involving every (33) 
state.
Not shown in figure \ref{delPezzo1}, but as held for $dP_0$, is that the choice of which bulk 4-cycle the 7-brane wraps allows us to couple the (37) 
states to any
given (33) state, independent of flavour.

The number of D7 branes is bound by the tadpole/non-abelian anomaly
cancellation as in the $dP_0$ case, which in this case reads:
\bea
m_4 &= &n_4+n_3-n_1-n_2+m_1 -m_2+m_3, \nonumber \\
m_5 & = & n_1-2n_2+n_4+m_2-m_3,
\nonumber \\
 m_6 &= & n_4-3n_1+2n_3+m_1-m_2
\eea

That is, given the number of D3 branes at each node $n_i$, the models
are determined by fixing also the number of D7 branes at the first
three nodes in the figure $m_{1,2,3}$. Solutions with $m_{i} \geq 0$
are physically relevant.

Similar to the $dP_0$ case the anomaly free combination of $U(1)$'s is:
\be
Q_{anomaly-free} = \sum_i \frac{Q_i}{n_i}
\ee

Once $\langle \Phi \rangle \neq 0$, the matter content is Higgsed back to the $dP_0$ quiver for energies $E \ll \langle \Phi \rangle$.
In principle there is no objection to $\Phi$ obtaining a vev, provided
that the mass of the  $Z'$ that
the vev would produce is beyond experimental bounds.
Indeed, in a realistic model it is necessary that one node of the quiver (the Higgs) is radiatively vevved during supersymmetry breaking.
It is not inplausible that this is not the only node that is vevved by the process of supersymmetry breaking.

      The $SU(2) \ti U(1)$ family symmetry,  allows us to engineer family symmetries that are less restrictive than the models
discussed in \ref{dP0sec}. Upon diagonalization, the mass squared matrix $MM^\dagger$ associated with the superpotential  (\ref{sp}) takes the form
$$
 \left( \begin{array}{ccc} M^{2} & 0 & 0 \\ 0 & m^{2} & 0 \\0 & 0 & 0 \end{array} \right).
$$
with $M\gg m$ for small values of the vev of $\Phi$, $\frac{\langle\Phi\rangle}{\Lambda}\ll 1$. This provides a
more realistic hierarchy of fermion
masses than
the $dP_0$ models.\footnote{The limit
$\frac{\langle\Phi\rangle}{\Lambda}\rightarrow 1$ corresponds to the $dP_{0}$
quiver, in this limit $m \rightarrow M$ restoring the $SU(3)$ symmetry of $dP_0$ .}. Further suppressed instanton
 contributions to Yukawa couplings have been
recently  computed for branes at singualrities models in \cite{iu}.

\subsubsection{Standard and Left-Right Symmetric $dP_1$ Models}

\begin{figure}
\twographs{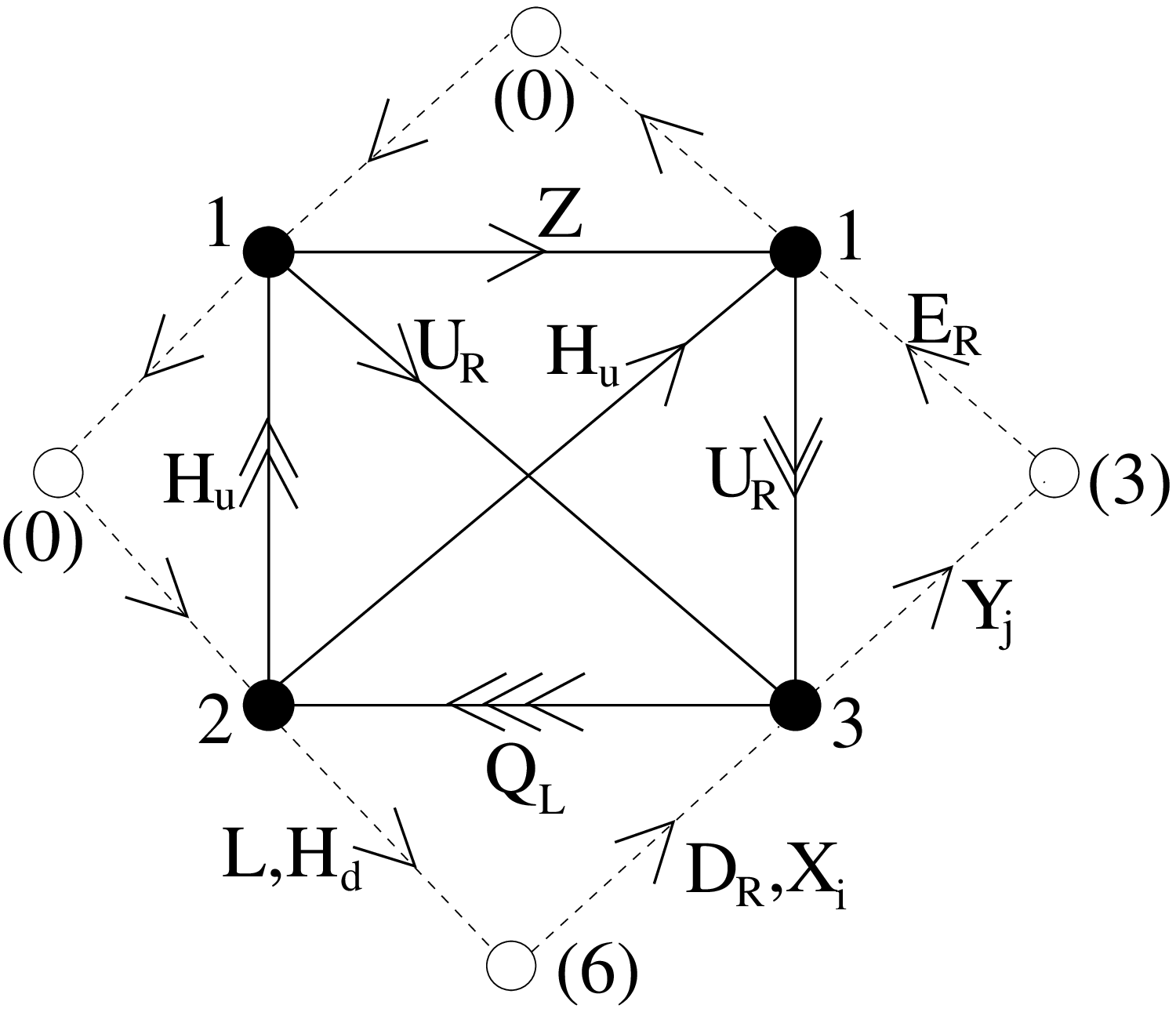}{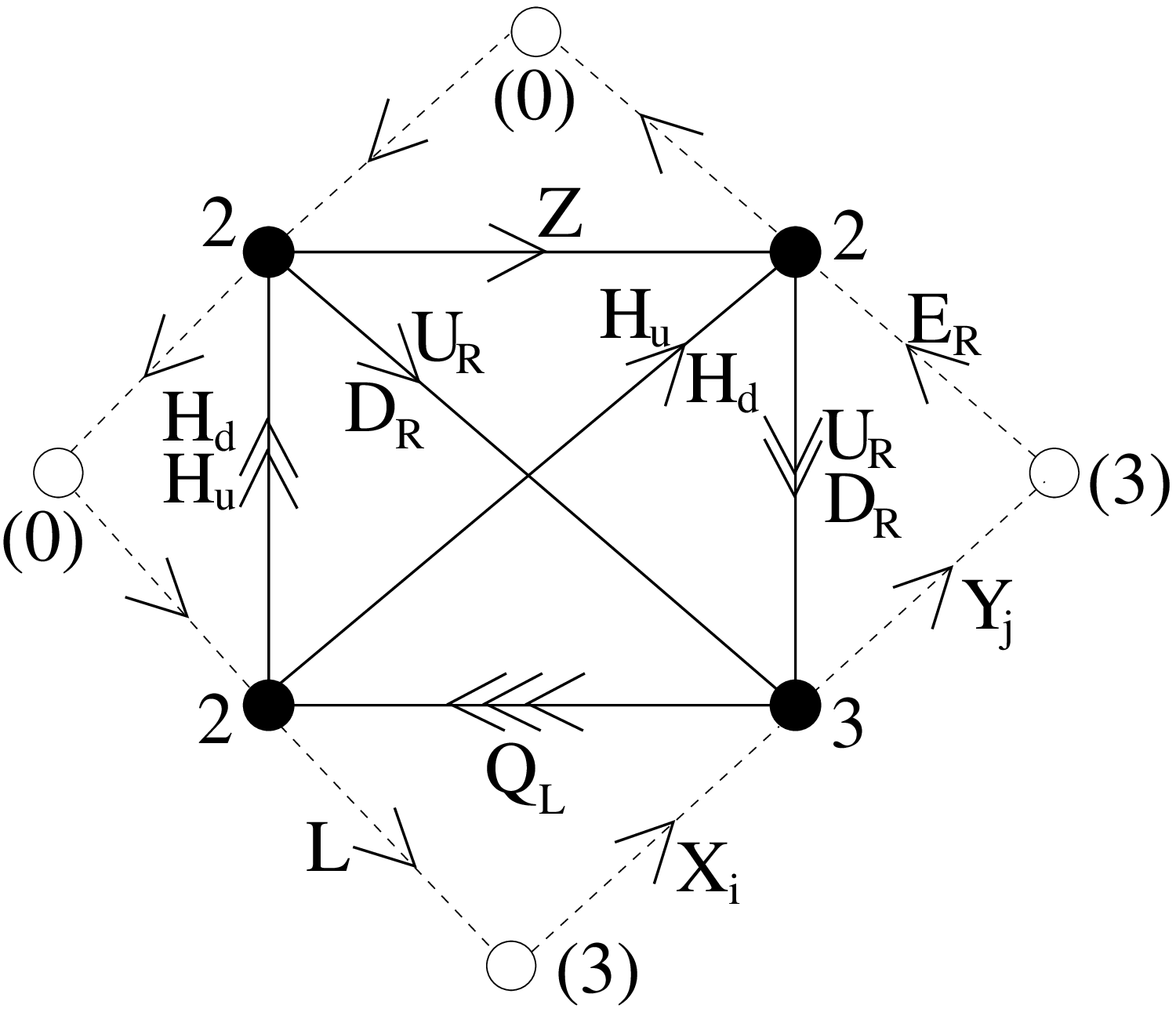}
\caption{A MSSM-like model and LR-symmetric model based on the del Pezzo 1 singularity. }
\label{dP1models}
\end{figure}

We can modify the $dP_0$ models of section \ref{dP0sec}  to obtain models with more realistic Yukawa couplings, with
a flavour symmetry of $SU(2) \ti U(1)$.
In figure \ref{dP1models} we show some quasi-realistic models based on the $dP_1$ singularity.

The first figure of \ref{dP1models} shows a quiver for generating an MSSM-like model.
In the non-compact limit, the anomaly-free gauge group of this model is $SU(3) \ti SU(2) \ti U(1)_Y \ti U(1)_Z$.
The additional $U(1)$ compared to the $dP_0$ models comes from the presence of the splitting of the $U(1)$
 node into two separate $U(1)$s, joined by the field $Z$. One of these $U(1)$s corresponds to hypercharge and the other
 to an additional $U(1)_Z$ under which different quark generations (in particular the $U_R$ fields) have different charges.

In the non-compact limit, $U(1)_Z$ is massless. However in a compact model it will acquire a mass through
the Green-Schwarz mechanism, provided all 2-cycles of $dP_1$ remain 2-cycles of the Calabi-Yau. As we shall see below,
$U(1)_Z$ will then acquire a mass of $M_{KK}$, the bulk KK scale of the compactification, and decouple from the low-energy physics.
In this case the gauge group returns to the case of $dP_0$ (with the addition of the neutral $Z$ field)
while the structure of Yukawa couplings is set by an $SU(2) \ti U(1)$ flavour symmetry.

The second figure of \ref{dP1models} shows a quiver for generating a Left-Right Symmetric model from the
$dP_1$ singularity. In this case, it is necessary to vev the $Z$ field in order to break the spectrum and gauge group back down
to that of the $dP_0$ case.
It may be asked why the $dP_1$ quiver is relevant at all, if it is necessary to vev it back down to the
$dP_0$ quiver. The collapse of the $dP_1$ quiver to the $dP_0$ quiver corresponds to the fact that
upon vevving the Z field, in the absence of SUSY breaking the $dP_1$ theory flows to the $dP_{0}$ theory  in the deep infrared.
By considering $dP_{1}$ models
in which SUSY breaking occurs well before the theory has evolved to the $dP_{0}$ theory,
\footnote{This can be achieved if $ \langle Z \rangle \ll M_{\rm{pl}}. $} one can obtain models in which the interactions are
different from theories at $dP_{0}$ quivers. In particular,
the flavour symmetry is that associated with the $dP_{1}$ geometry,
while the low energy matter content is that associated with a $dP_{0}$  quiver.

Some comments are appropriate about the relationship of the vevved $dP_1$ quiver and the $dP_0$ models. By blowing up a 2-cycle in the
$dP_1$ geometry, the geometry of the actual singularity reduces to that of $dP_0$. It may therefore seem more appropriate to describe the
singularity as $dP_0$. However, if the vev of the blow up field is substantially sub-stringy, and so far away from the geometric regime, this vev is most
straightforwardly viewed as a perturbation on $dP_1$ within field theory
rather than as $dP_0$ with a nearby resolved cycle in the geometry. This latter viewpoint would be more appropriate if the cycle was resolved with a string/Planck
vev taking it all the way into the bulk geometric regime.

Further models can be constructed based on $dP_1$: it is in principle possible to consider a node in the quiver without any branes.
This immediately fixes  the modulus corresponding to fractional branes moving out of the singularity
and avoids
reducing the gauge group to factors smaller than the Standard Model. However, it is not clear how the correct hypercharge assignments will emerge
in this case (and this modulus is expected to be fixed by soft supersymmetry breaking terms).
We could also
orientifold these models opening the possibility of reducing the number of extra doublets as studied in the second reference of \cite{alday}, and
also potentially introducing symmetric and antisymmetric representations. 
 
\subsection{Higher del Pezzos}

Using the methodology of adding nodes and vevving them, it is easy to extend any of the above models to
any of the higher del Pezzos. The motivation for doing so however weakens as the del Pezzo rank is increased: the
extent of flavour symmetries decrease, and the amount of vevving required to reduce to the Standard Model spectrum makes
the models increasingly baroque, without substantially improving the phenomenology.

\section{Global Embeddings}
\label{sec3}

\subsection{Local/Global Mixing}

To study supersymmetry breaking in a controlled fashion, it is necessary to embed the above local constructions into
global models in which the moduli are stabilised and supersymmetry is broken.
Recently Blumenhagen, Moster and Plauschinn have
emphasised a difficulty with combining realistic chiral matter sectors with
moduli stabilisation \cite{07113389}.
In the IIB context the issue can be summarised as follows. In the IIB context,
moduli stabilisation techniques involve 3-form fluxes stabilising the complex structure moduli and
non-perturbative effects stabilising the K\"ahler moduli. The typical moduli effective action takes the form
\bea
K & = & - 2 \ln \left( \mc{V} + \frac{\xi (S + \bar{S})^{3/2}}{2} \right) -
\ln \left( i \int \Omega \wedge \bar{\Omega}(U, \bar{U}) \right) - \ln (S + \bar{S}), \\
W & = & \int G_3 \wedge \Omega(U) + \sum A_i(U) e^{-a_i T_i}.
\eea
Here $T_i$ are K\"ahler moduli, $U_j$ complex structure moduli and $S$ is the dilaton.
$\xi$ is a numerical factor representing the $\alpha'^3$ correction to the K\"ahler potential. After flux stabilisation, the effective
theory for the K\"ahler moduli is
\bea
K & = & - 2 \ln \left( \mc{V} + \frac{\xi'}{g_s^{3/2}} \right), \\
W & = & W_0 + \sum A_i e^{-a_i T_i}.
\eea
The justification for integrating out the $U$ moduli is essentially
the factorised form of the K\"ahler potential and the lack of cross-couplings between $U$ and $T$ fields.
For a recent discussion of the consistency of integrating out
moduli in supergravity, see \cite{08064364}. 
The presence of a `bare' instanton superpotential $e^{-a_i T_i}$ requires the instanton to have only two fermionic zero modes
and the modulus $T$ to be uncharged. This occurs for example for instantons wrapping rigid blow-up cycle, where there are no
massless adjoint degrees of freedom.

By definition however there are branes and chiral fermions on the cycles supporting the MSSM.
Instantons wrapping the same cycle as any MSSM brane have a non-zero intersection number with
such branes, giving rise to extra fermionic zero modes. This forbids the bare term $e^{-a T_{MSSM}}$
from appearing in the superpotential and requires it to be instead dressed with matter fields.
Equivalently, in brane constructions the chiral nature of the MSSM implies the
existence of anomalous U(1)s under which moduli are charged,
$\delta_{\lambda} T_{MSSM} = T_{MSSM} + i Q_T \lambda$. For all such moduli the term $e^{-a T_{MSSM}}$ is gauge-variant
and cannot appear bare in the superpotential.

The consequence is that if $T_{MSSM}$ appears in the superpotential, it can only do so in the
gauge-invariant form
$$
\left( \prod_i \Phi_{hidden,i} \right) \left( \prod_j \Phi_{MSSM,j} \right) e^{-a T_{MSSM}}.
$$
However $\Phi_{MSSM}$ does not acquire a vev,\footnote{The Higgs vev is too small to be relevant for moduli stabilisation.}
and so there is no
non-perturbative superpotential available to stabilise $T_{MSSM}$.

There are two basic approaches to this problem.
One could suppose as in \cite{08051029} that the
MSSM cycle size is stabilised by loop (worldsheet or spacetime) corrections
to the K\"ahler potential. The difficulty here is that such corrections are hard to calculate in a controlled way,
and it is not easy to ensure the cycle is stabilised in the geometric regime.

Reference \cite{07113389} suggested aiming to stabilise the Standard Model cycle
using D-terms for anomalous U(1)s. Such D-terms take the form
$$
D_a^2 \sim \sum_i \left( \vert \Phi \vert^2 - \xi \right)^2,
$$
where $\xi$ is the moduli-dependent Fayet-Iliopoulos term
$\xi = (\partial_{V_a} K) \vert_{{V_a}=0}$
and $V^a$ is
the $U(1)$ vector multiplet.
In the geometric regime the FI term can be written as $\xi \sim \int J \wedge F$.
If $\Phi_{MSSM}$ is forced to vanish, these K\"ahler moduli are stabilised by
$\xi = 0$.

In \cite{07113389} a toy model was studied where D-terms constrained the `Standard Model' cycle
to finite size while another unrelated cycle collapsed
to the edge of the K\"ahler cone.
A general disadvantage of using the geometric expressions for
D-terms to stabilise moduli is that the FI term has a tendency to drive cycles to collapse, i.e. to the
boundary of the K\"ahler cone. However in this regime the FI term will be modified
by corrections to the K\"ahler potential. Furthermore, branes that were
originally BPS in the geometric regime may
become unstable and decay to a new set of stable branes.
It is instead necessary to use the BPS brane states associated to the collapsed geometry, but it is not
easy to follow this transition through.

An attractive feature of models of branes at singularities is that they
allow a promising possible resolution of this tension. As described in section \ref{sec2}, the
stable BPS branes at the singularity are known and their matter content and interactions
are encoded in the quiver/dimer diagrams. We have also seen that
realistic matter spectra occur rather naturally in this framework.
The FI terms for the anomalous U(1)s correspond to
vevs for the blow-up modes that resolve the singularity.
Requiring vanishing FI terms stabilises the blow-up moduli at the singularity.
Even though this is on the edge of the K\"ahler cone, we have excellent
model-building control here as we know the appropriate set of fractional branes
that apply at the singularity. For the $\mbb{C}^3/\mbb{Z}_3$ singularity the geometry is even simpler
and is a very simple orbifold singularity.

We note that strictly speaking what is fixed by the D-term is a
combination of the matter field vevs and the FI term, as a
non-zero $\langle \xi \rangle$ can always be cancelled by a non-zero $\langle \Phi \rangle$.
However, after supersymmetry breaking the
presence of soft scalar masses for the matter fields $\Phi$ will lift this degeneracy.
As $\Phi=0$ is a point of enhanced symmetry, we consider it a reasonable assumption that after supersymmetry breaking
soft scalar masses will fix $\Phi =0$. It would however clearly be nice to verify this assumption
in a full model.

For clarity let us enumerate the steps required (also see \cite{compact} for similar ideas):
\begin{enumerate}
\item
The D-term generates a potential
\be
V_D \sim (\sum_i \vert \Phi_{MSSM} \vert^2 - \xi)^2
\ee
that fixes a combination of the matter fields $\Phi_{MSSM}$ and the unfixed K\"ahler moduli (in $\xi$) but leaves an overall
flat direction.
\item
The presence of soft scalar masses for the matter fields, induced after supersymmetry breaking, will lift this degeneracy.
\item
If the soft masses are positive in an expansion about $\Phi = 0$ then the full minimum of the potential will be at $\Phi = 0$
and $\xi =0$ thereby fixing the K\"ahler moduli at the singularity. While it is not possible to determine the sign of the soft masses
without a full study of supersymmetry breaking, this sign does only represent a discrete parameter.

We also note that in a phenomenological model the positive mass is phenomenologically necessary (except for the Higgs scalars) in order
to avoid charge and colour breaking minima. Thus in models in phenomenologically realistic supersymmetry breaking the blow-up moduli
will be fixed at the singularity.

\end{enumerate}

We now flesh out this picture and describe the requirements
on the global geometry in order to realise this embedding.

\subsection{Towards Fully Global Models}

We wish to embed the above local models into a  bulk that stabilises moduli and breaks
supersymmetry. We will base the bulk on the LARGE volume method of moduli stabilisation \cite{hepth0502058, hepth0505076}.
Under rather general conditions, analysed in most detail in \cite{08051029},
this stabilises moduli and gives controlled and dynamical low-energy supersymmetry breaking.
The principal characteristic of this model is the exponentially large volume. The stabilisation and
phenomenology of these models have been studied in
\cite{hepth0505076, hepth0605141, hepth0610129, 07040737, 07070105, 07113389, 08041248, 08062667, 08103329}.
General features of these models are
\begin{enumerate}
\item
A `Swiss-cheese' geometry, with a large bulk volume and several small blow-up cycles.
\item
A gravitino mass
\be
m_{3/2} = \frac{W_0 M_P}{\mc{V}}.
\ee
\item
Supersymmetry dominantly broken by the volume modulus in an approximately no-scale fashion.
\end{enumerate}

\begin{figure}
\begin{center}
\includegraphics[width=8cm]{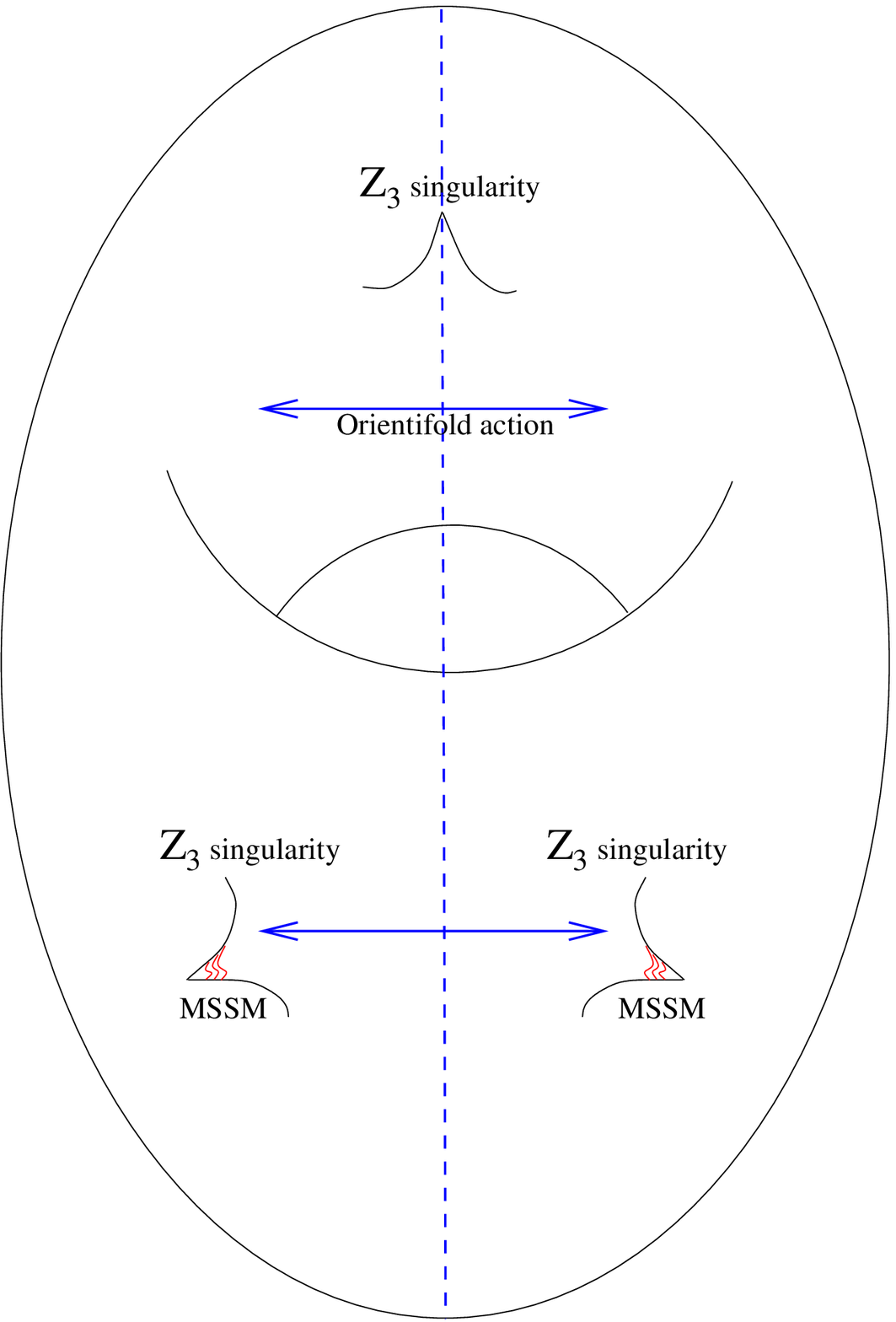}
\caption{A schematic of the required Calabi-Yau properties to generate a model with a realistic matter sector, full moduli
stabilisation and controlled hierarchically small dynamical supersymmetry breaking.
\label{GlobalModel}}
\end{center}
\end{figure}
In figure \ref{GlobalModel} we provide a schematic of the geometry required for the minimal global embedding.
The minimal geometry consists of a Calabi-Yau with 4 K\"ahler moduli, of which one ($\tau_b$) controls the overall volume and
three ($\tau_{s,i}$) are blow-up modes resolving singularities. Such shrinkable 4-cycles are del Pezzo surfaces and thus rigid:
a brane wrapping such a surface has no adjoint matter.
For simplicity we assume the local
geometry around each blow-up mode to be the cone over
$dP_0$, i.e. the geometry of the $\mbb{Z}_3$ singularity.
The volume can be written in the `Swiss-cheese' form
\be
\mc{V} = \tau_b^{3/2} - \tau_{s,1}^{3/2} - \tau_{s,2}^{3/2} - \tau_{s,3}^{3/2}.
\ee
In this example there are a total of four K\"ahler moduli. This is a minimal requirement: one to control the overall volume,
one to have non-perturbative effects, and two to be exchanged by the orientifold. This framework can be trivially generalised
to models with more K\"ahler moduli. The `Swiss-Cheese' form of the volume is known to exist for models with 2 ($\mbb{P}^4_{[1,1,1,6,9]}$),
3 ($\mbb{P}^4_{[1,3,3,3,5]}$), and 5 K\"ahler moduli ($\mbb{P}^4_{[1,1,3,10,15]}$) (the last two are described in \cite{07113389}).
Further examples are also discussed in \cite{08053361}.

We require that an orientifold action leading to O3/O7 planes is well-defined on the Calabi-Yau.
An orientifold requires invariance under the action of $\Omega \sigma (-1)^F$, where $\Omega$ is world-sheet parity, $F$ is world-sheet
fermion number, and $\sigma$ is a $\mbb{Z}_2$ involution of the Calabi-Yau. The involution $\sigma$ acts on the
various 4-cycles and we use $h_{1,1}^{+}$ ($h_{1,1}^{-}$) to denote the number of 4-cycles with positive (negative) parity
under $\sigma$. The moduli content after the orientifold is given by
\bea
h_{1,1}^{+} & & T_{\Sigma} + i C_4 \\
h_{1,1}^{-} & & B_2 + i C_2.
\eea
We require that under the involution the large 4-cycle, $\tau_b$, and one of the
small blow-up cycles, for definiteness $\tau_{s,1}$, is taken to itself.
In the case that an orientifold plane wraps this cycle we place D7 branes
on top of the O7-plane, cancelling the local tadpoles and generating an $SO(8)$ gauge group.
As the cycle is a blow-up cycle it is rigid and a brane wrapping it
carries no adjoint matter and no additional fermionic zero modes.
The $SO(8)$ gauge group therefore undergoes gaugino condensation and
generates a non-perturbative superpotential in $\tau_1$.
If no orientifold plane wraps the $\tau_1$ cycle, then
D3-instantons will generate a non-perturbative superpotential for $\tau_1$.
Such a nonperturbative superpotential for $\tau_1$ is necessary to obtain the LARGE volume stabilisation.

We also require the involution to exchange the remaining two small cycles, $\tau_{s,2} \leftrightarrow \tau_{s,3}$.
This will ensure the local geometry near these singularities is that of a pure Calabi-Yau singularity.
The orientifold action simply relates the physics at one singularity to
that at the other. Orientifolded singularities may also be interesting for model-building, but for simplicity we do not consider them here.
At these singularities we introduce
one of the models of section \ref{sec2}, giving a realisation of a chiral matter sector
containing the Standard Model matter content. These models cancel all local tadpoles
and only leave a bulk D7 tadpole and a D3 tadpole.
For the trinification model, there is no D7 tadpole and the only tadpole to be cancelled is the D3 tadpole.

As well as the construction of an appropriate Calabi-Yau
- for example using the models of \cite{08053361} - many other conditions must be satisfied to build a fully consistent global model.
These include the cancellation of all RR tadpoles, the specification of explicit 3-form flux quantum numbers
and the solution of the flux equations $D_U W = 0$,
and cancellation of Freed-Witten anomalies between fluxes and branes.
Many of these conditions are more mathematical in nature, and do not seem to have direct effect on the phenomenology
or the pattern of supersymmetry breaking.
While the explicit construction of such global models is important, it is beyond the scope of this paper
(see \cite{esole} for a recent
detailed discussion of consistency conditions for
IIB model building).

\subsection{Effective Field Theory Near the Singularity}

For a given compactification we usually know that an effective field theory can be used in the regime where the moduli are
larger than the string scale $l_s$. Starting in this regime, the effective field theory ceases to be valid when we approach the boundary of the 
K\"ahler cone in which one of the moduli collapses to zero size. However, string theory is known to behave properly at singularities. Therefore there 
exists an effective field
theory description close to the singularity,  in the regime for which the overall size of the
corresponding blow-up mode is much smaller than the string scale $l_s$. We therefore have two different effective field theories depending on which 
regime we are considering the blow-up mode. Most of the studies have been done in the large modulus regime corresponding to magnetised D7 branes 
models.
We collect here the main expressions for the effective field theory in the vicinity of the singularity. For concreteness we will assume that the
Standard Model cycle of size $T_{SM}$ is close to the singularity whereas the 4-cycle of size $T_s$
providing the non-perturbative superpotential is in the large modulus regime.

As usual we need expressions for the gauge kinetic functions $f$ , the
superpotential $W$ and the K\"ahler potential $K$. The gauge kinetic function
 in the magnetised D7 brane regime look like $f=T_s+\alpha S$, with $T_s$ the 4-cycle modulus, $S$ the dilaton and $\alpha$ a flux dependent 
coefficient.
Close to the singularity the gauge kinetic function takes the form $f= S+ \beta T_{SM}$, with $\beta $ a loop correction parameter.

For the superpotential, as usual RR and NS-NS fluxes give rise to  a constant superpotential $W_0$ (after stabilising complex structure and dilaton
moduli), non-perturbative effects give rise to the standard  $e^{-aT_s}$ term. Yukawa couplings differ substantially whether the standard model is at 
the
singularity compared to the large blow-up limit. In \cite{08023391,08070789} it was shown that the Yukawa couplings in a blown-up $P_2$ vanish 
identically, however it is known that at the singularity the Yukawa couplings are generally non-vanishing and determined by the structure of the 
quiver/dimer diagram as we have discussed above.
\be
W  =  W_0 + A e^{-a T_s} + Y_{ijk} C^i C^j C^k.
\ee

$Y_{ijk}$ is singularity dependent (for $dP_0$ it is $\epsilon_{ijk}$).
The K\"ahler potential is more difficult to determine and takes the form:
\bea
\mc{K} & = & - 2 \ln ((T_b + \bar{T}_b)^{3/2} - (T_s + \bar{T}_s)^{3/2} + \xi) + \frac{(T_{SM} + \bar{T}_{SM} - q V)^2}{\mc{V}} \\
& & + \frac{(B_2 + \bar{B}_2 - q' V')^2}{\mc{V}} + \frac{C^i \bar{C}^i}{\mc{V}^{2/3}} \left( 1 + \mc{O}((T_{SM} + \bar{T}_{SM})^\lambda + \ldots 
\right),
\eea
with $\lambda > 0$. Here $C^i$ are chiral matter fields, $T_{SM}$ and $B_2$ the local moduli of the singularity, and $q$ and $q'$ the charges of the 
moduli
under the two anomalous $U(1)$s of the $dP_0$ singularity. $V$ and $V'$ are the vector multiplets of these $U(1)$s.
The kinetic term factors of $\frac{1}{\mc{V}}$ for the local modulus $T_{SM}$ and $\frac{1}{\mc{V}^{2/3}}$ for the local matter fields will be
justified as follows.

Let us start with the matter fields for which we follow an argument similar to \cite{ccq}.
The volume dependence of $\mc{V}^{2/3}$ is equivalent to the statement that the physical Yukawa couplings are local
and do not depend on the overall volume. This follows from the expression for the physical Yukawas $\hat{Y}_{\alpha \beta \gamma}$,
\be
\hat{Y}_{\alpha \beta \gamma} = e^{\mc{K}/2} \frac{Y_{\alpha \beta \gamma}}{\sqrt{K_{\alpha} K_{\beta} K_{\gamma}}}.
\ee
As the superpotential Yukawas cannot depend on the volume moduli due to holomorphy, the dependence of the K\"ahler
metric on the volume is
fixed by the requirement that $\hat{Y}_{\alpha \beta \gamma}$ does not depend on the bulk volume.
The absence of any leading order dependence on $T_{SM}$ also follows from the finiteness of the Yukawas: at the singularity
$T_{SM} = 0$, while the physical Yukawas are finite and non-zero.

Let us now consider the
 volume scaling, $K_{B_2} \sim K_{T_{SM}} \sim \frac{M_P}{\mc{V}}$. There are three ways to understand the volume scaling
of this.
\begin{enumerate}
\item
This is the volume scaling that holds in the geometric regime (e.g. for $T_s$). Collapsing to the singularity is a local effect and will not
affect the power of volume that appears.
\item
As we will see in the next subsection, the mass of an anomalous U(1) is given by $m_{U(1)}^2 = M_P^2 K''(T_{SM} + \bar{T}_{SM})$. As we calculate
$m_{U(1)}^2 = M_s^2 = \frac{M_P^2}{\mc{V}}$, this fixes the volume scaling of $K(T_{SM} + \bar{T}_{SM})$ to be $1/\mc{V}$.

\item
We can imagine moving to a point in field space where $T_{SM}$ is resolved to finite size comparable to the string scale but where none of the matter
fields have vevs.
This configuration breaks supersymmetry in a hard fashion. The vacuum energy comes from a non-supersymmetric brane configuration and will be
$V \sim M_{string}^4 \sim \frac{M_P^4}{\mc{V}^2}$. As this energy is associated with the D-term this implies that
$$
V_D = \frac{M_P^4}{2} \hbox{Re}(f_a)^{-1} \left( Q \partial_{T_{SM}} K \right)^2 \sim \frac{M_P^4}{\mc{V}^2}.
$$
It follows that if $T_{SM}$ measures the size of the resolution in string units, and is thus
$\mc{O}(1)$ for a resolving geometry of characteristic radius
$\sqrt{\alpha'}$ and characteristic energy $V \sim M_{string}^4$,
$$
K(T_{SM}, \bar{T}_{SM}) = \frac{(T_{SM} + \bar{T}_{SM} + qV)^2}{\mc{V}}.
$$
The shift symmetry is associated with the axionic nature of $\hbox{Im}(T)$.
\end{enumerate}

Given this geometry, the LARGE volume stabilisation mechanism of \cite{hepth0502058} gives rise to
both moduli stabilisation and dynamical supersymmetry breaking, with the volume stabilised at an exponentially large value.
For this to occur we require that the compact space $X$ has more complex
structure moduli than K\"ahler moduli, $h^{2,1}(X) > h^{1,1}(X)$. This condition is due to a requirement on the sign
of the $\alpha'^3$ correction to the K\"ahler potential, which depends on the sign of the Euler number of the Calabi-Yau.

Finally we would like to emphasise the following important feature that can be confusing. One of the properties of
local models is that,
contrary to common lore, not all the moduli couple with gravitational strength interactions. It is clear that the matter fields
living in a local cycle couple to the volume modulus with gravitational strength interactions since their couplings probe the whole manifold.
 The same applies to couplings to other moduli. However the interaction of
matter fields on a local cycle with the moduli controlling the size of that cycle are only
  suppressed by the string scale and not by the Planck scale.
  Explicit calculations illustrating this fact can be found  in reference \cite{cqastro}. This result
will play a important role when we discuss soft supersymmetry breaking.

\subsection{Phenomenological Features}

\subsubsection{Hyperweak Forces}

In local models of branes at singularities, the global D7 branes provide global symmetries of the model. However once the corresponding model is 
embedded in
a compact model, the symmetries induced by these D7 branes will be
gauged with an inverse gauge coupling of order the size of the bulk cycle.
As emphasised in reference \cite{global}, the D7 branes could play an important phenomenological role. Since they probe the global structure of the 
extra dimensions, they naturally wrap the exponentially large 4-cycle and the corresponding gauge coupling will be exponentially small ($g^{-2}\sim 
\tau_b\sim {\cal V}^{2/3}$) (see \cite{hepth0111269} for related work).
 Therefore the Standard Model states coming from
 D3-D7 states will be charged under these extra symmetries. In some of the examples above
 these particles are the leptons but there are also models with the right handed quarks corresponding to D3-D7 states.
 The existence of these remarkably weak interactions could be considered as an interesting way to test some of these models. Estimates for the masses 
of the extra gauge bosons and their phenomenological implications are discussed in \cite{global}.

Here we note that such hyperweak gauge bosons may be relevant for explaining
the ghost muon anomaly recently seen at the CDF detector at the TeVatron \cite{CDFpaper}.
Di-muon events with tight identification within the inner silicon vertex (SVX) trackers (and small impact parameters $\sim \mc{O}(0.1 \hbox{cm})$)
are well-described by Standard Model and detector effects, giving a $\sigma_{b \to \mu, \bar{b} \to \mu}$ cross-section in good agreement with
NLO theoretical predictions.
However there are also a large number of so-called `ghost' muon events, which register on loose SVX cuts,
with large muon impact parameters $\mc{O}(1 \hbox{cm})$.
These ghost muons have larger additional nearby particle multiplicity than for muons from known QCD events
and have a broad peak in their invariant mass spectrum at $\mc{O}(1)\, \hbox{GeV}$.
CDF is currently unable to account
for the presence of these ghost muons with their large impact parameters using their current understanding of Standard Model backgrounds and detector 
performance.
While this may be due to experimental subtleties in understanding backgrounds, particle reconstruction or other detector effects
- and most experimental anomalies do in time go away - there is however the possibility that the ghost muons are due to the decay
of a previously unknown long lived particle which we will henceforth denote $U$.

If such an interpretation is correct, CDF's fit to the distribution of impact parameters gives a lab lifetime
$\tau \sim 2 \ti 10^{-11}s$ for this new particle. The large numbers of events with ghost
muons (73000 events compared to 195000 for $b \to \mu, \bar{b} \to \mu$ for $2 \hbox{fb}^{-1}$) suggests this $U$ is
produced abundantly and with relatively large cross-section, even if it is possible to attribute
an $\mc{O}(1)$ fraction of ghost events to non-exotic physics.
This argues against interpretations in which $U$ is accessed only through a very heavy portal, e.g. supersymmetric
cascade decays of the gluino. Such processes would have difficulty producing sufficient $U$ events while remaining
consistent with null results for new high $P_T$ physics searches at the TeVatron.

Light weakly coupled particles are naturally long lived.
A hyper-weak gauge boson can be naturally light, with a mass
$m_{_U} \sim v  g_{_U} $ where $v$ is the vev of the $U$ Higgs.
For weak-scale vevs (as for the SM Higgs) the $U$ boson mass is significantly suppressed compared to $M_W$ by the small coupling.
For direct two-body decay, on dimensional grounds the $U$ boson lifetime would be
\be
\label{Ulifetime}
\Gamma \sim \frac{g_{_U}^2 m_{_U}}{4 \pi}, \qquad \tau \sim \frac{4 \pi}{g_{_U}^2 m_{_U}} \sim \left( \frac{1 \hbox{GeV}}{m_{_U}} \right)
\left( \frac{10^{-8}}{g_{_U}^2} \right) 10^{-16}s.
\ee
The lifetime is automatically enhanced compared to e.g. QCD timescales by the low mass and weak coupling.

In order for $U$ to be abundantly produced at a hadron collider, quarks must be charged under it.
A light $U$ can be produced either directly through $q \bar{q}$ annihilation
or alternatively as initial state (ISR)  or final state (FSR) radiation in any regular QCD process. For either ISR or FSR
emission probabilities of a light $U$ boson are logarithmically enhanced by standard Sudakov factors.
If $m_U \sim 1 \hbox{GeV} \sim \Lambda_{QCD}$ then on dimensional grounds we expect $U$ boson emission from the initial or final state
to be suppressed compared to QCD emission by $\alpha_U/\alpha_{QCD}$. While this number is small, the overall number of events involving
$U$ emission may still be very large given the colossal size of the QCD background at a hadron collider.

Supposing a new $U$ particle to be responsible for the ghost muons, the CDF dimuon invariant mass spectrum
shows no sign of a sharp narrow muon resonance
associated to 2-body $U \to \mu^{+} \mu^{-}$ decay.
Instead a broad plateau is seen at invariant energies $m_{\mu \mu} \sim 1 \hbox{GeV}$. While the presence of
(possibly multiple) muons in $U$ decay implies muons must exist as $U$ decay products,
 the absence of a resonant peak implies the 2-body decay $U \to \mu^{+} \mu^{-}$
should have small to negligible branching ratio.
Interpreting $U$ as a gauge boson, this suggest that leptons (or at least muons) should not have gauge couplings to
$U$.

For $m_{_U}\sim 1 \hbox{GeV}$ the $U$ lifetime (\ref{Ulifetime}) is still too short to account for the large impact factors of ghost muons.
However if $U$ is light, with quarks charged under it, then $U$ may have no kinematically accessible on-shell 2-body decay modes.
For example, it is trivially true that for $m_{_U} < m_{\pi}$, no hadronic final states at all are accessible.
This effect may be enhanced if $U$ has family-dependent couplings so that - for example - $U$ decays must necessarily involve second generation
quarks. In any event, $U$ is obliged to decay either through loops or off-shell hadrons which in turn decay to the observed particles.
The additional suppression from such decays can then bring the $U$ lifetime to the $\mc{O}(10)\, \hbox{ps}$ range necessary to account for the ghost
muon anomaly. If such higher-order decays involve the weak interaction (or have as products short-lived hadrons with significant
(semi)-leptonic decay modes) then muons can be produced in $U$ decays without a resonant muon peak.

In summary, we propose that hyperweak forces could be responsible for the CDF ghost muon anomaly in the following way:
\begin{enumerate}
\item
There exists a light hyperweak gauge boson $U$ of mass $m_{_U} \sim 1 \hbox{GeV}$ under which (some) quarks are charged
but muons are not charged. The $U$ gauge coupling is substantially (possibly exponentially) weaker than
the Standard Model gauge couplings.
\item $U$ is produced relatively abundantly in high energy $p \bar{p}$ collisions either through direct production or
through initial/final state radiation off quarks in any vanilla QCD event. In particular, $U$ does not require
TeV-scale new physics as a portal for its production.
\item $U$ has no direct $\mu \mu$ coupling and
kinematics forbids the direct decay of $U$ into 2-body hadron final states. $U$ is instead obliged to decay through
off-shell hadronic final states. Kinematics require some of these off-shell processes to involve the weak interaction
resulting in muons as a decay product.
\item
The combination of the weakness of the initial gauge coupling, the lightness of $U$ and the absence of direct kinematically
accessible final states leads to a displaced $U$ vertex and the macroscopic muon impact parameters observed by CDF.
\end{enumerate}

Notice that this proposal to explain the CDF data can be  considered independently of string theory.
While we have assumed $U$ to be a gauge boson motivated by the D3-D7 brane models considered in this paper, the same basic phenomenological
approach could apply for any bulk state with very weak couplings to quarks and no kinematically accessible 2-body decays.
In the LARGE volume scenario  a mass $m_{_U}\sim 1 \hbox{GeV}$ (and a Standard Model scale $vev$) implies a volume of order $\mc{V}\sim 10^8$
which would correspond to $M_s\sim 10^{14} \hbox{GeV}$.
It remains to be seen if with these numbers it is possible to obtain the proper $U$ lifetime.

The above is clearly a phenomenological scenario and not yet a full model. More detailed model-building is necessary for a full analysis of the
merits and further phenomenological and cosmological implications of the approach outlined. However, it provides a proposal for how such additional 
hyper-weak
$U(1)$s could manifest themselves in collider phenomenology with possible relevance for the current CDF ghost muon anomaly.

\subsubsection{Masses of anomalous and non-anomalous U(1)s}

It is well known that in brane models masses can be generated through the Green-Schwarz mechanism for U(1) gauge bosons that are non-anomalous.
For global brane models, the masses of such U(1) bosons are comparable to those of anomalous U(1) bosons and are close to the string scale.
In local models, the situation is more subtle as the non-anomalous U(1) bosons are massless in the non-compact limit and only acquire
masses on compactification. For large compact volumes the U(1) masses are hierarchically lower than the string scale. Our purpose here is
to compute the mass scale of such U(1)s.

The relevant Lagrangian is
\be
\mc{L} = - H_{\mu \nu \rho} H^{\mu \nu \rho} - \frac{1}{4g^2} F_{\mu \nu} F^{\mu \nu} - m D_2 \wedge F_2.
\ee
On dualising $D_2$ to a pseudoscalar this generates an explicit mass term $m^2 A_\mu A^\mu$ for the gauge boson.
We consider a model of D7 branes wrapping a local (collapsed) 4-cycle $X$ and use dimensional reduction to derive the
above Lagrangian. The $D_2 \wedge F_2$ term descends from
$$
\frac{2 \pi}{(2 \pi \sqrt{\alpha'})^8} \int \sum C_q \wedge e^{2 \pi \alpha' F} \to \frac{1}{2 \pi (2 \pi \sqrt{\alpha'})^4} \int_{\mbb{M}_4 \ti X} 
F_2 \wedge C_4 \wedge F_2.
$$
We can decompose $C_4$ as
\be
C_4 = D_2^i(x) \wedge \omega^i_2(y) + a^i(x) \tilde{\omega}^i_4(y).
\ee
Here $\omega_2$ is a harmonic 2-form with $\frac{1}{(2 \pi \sqrt{\alpha'})^2} \int_{\Sigma_j} \omega_2^i = \delta^i_j$.
Self-duality of $F_5 = dC_4$ relates $D_2^i$ and $a^i$. Reduction of $C_4$ along the brane gives a coupling
$$
\frac{Q_i}{2 \pi ( 2 \pi \sqrt{\alpha'})^2} \int D_2^i \wedge F_2,
$$
where $D_2$ is dimensionless. $Q_i = \int_X \omega_2 \wedge F_2$ is simply a number which is unimportant for our purposes.
Using $(2 \pi \sqrt{\alpha'})^{-2} \sim M_s^2 \sim \frac{M_P^2}{\mc{V}}$, we can write the $D \wedge F$ coupling as
$$
\frac{M_P^2}{\mc{V}} \int D_2 \wedge F_2 = \frac{M_P}{\mc{V}} \int (M_P D_2) \wedge F_2.
$$
To compute the mass, we also need the normalisation of the kinetic term $H_{\mu \nu \rho} H^{\mu \nu \rho}$. This term descends from
$$
\frac{1}{(2 \pi \sqrt{\alpha'})^8} \int dC_4 \wedge * dC^4 = \frac{1}{(2 \pi \sqrt{\alpha'})^8} \int d D_2^i \wedge * d D_2^j
\int \omega_2^i \wedge (* \omega_2)^j.
$$
If $t^i$ represents a 2-cycle size with $\mc{V} = k_{ijk}t^i t^j t^k$, then \cite{CandelasdelaOssa}
\bea
\frac{1}{(2 \pi \sqrt{\alpha'})^6} \int \omega_2^i \wedge * \omega_2^j & = & k_{ijk} t^k - \frac{3 (k_{ipq} t^p t^q) (k_{irs} t^r t^s)}{2 \mc{V}} \\
& \equiv & \mc{K}_{ij} - \frac{3 \mc{K}_i \mc{K}_j}{2 \mc{V}}.
\eea
This gives
\be
\frac{1}{(2 \pi \sqrt{\alpha'})^2} \int d D_2^i \wedge (* d D_2)^j \left( \mc{K}_{ij} - \frac{3 \mc{K}_i \mc{K}_j}{2 \mc{V}} \right).
\ee
Using $d D_2^i \wedge (* d D_2)^j = \sqrt{g} H_{\mu \nu \rho}^i H^{j, \mu \nu \rho}$, we obtain
$$
\frac{1}{\mc{V}} \int \sqrt{g} (M_P H_{\mu \nu \rho}) (M_P H^{\mu \nu \rho} ) \left( \mc{K}_{ij} - \frac{3 \mc{K}_i \mc{K}_j}{2 \mc{V}} \right).
$$

For anomalous $U(1)$s, the dual 4-cycle is compact in the local geometry and $\mc{K}_{ij} \sim \mc{O}(1)$. For non-anomalous $U(1)$s, the dual
4-cycle is non-compact and $\mc{K}_{\alpha} \sim \mc{V}^{2/3}$. The normalisation of the kinetic terms are thererfore
\bea
\textrm{Anomalous U(1)s} & \qquad & \int \sqrt{g} \left( \frac{M_P H^i_{\mu \nu \rho}}{\sqrt{\mc{V}}} \right) \left( \frac{M_P H^{j,\mu \nu 
\rho}}{\sqrt{\mc{V}}} \right) \\
\textrm{Non-anomalous U(1)s} & \qquad & \int \sqrt{g} \left( \frac{M_P H^i_{\mu \nu \rho}}{\mc{V}^{1/3}} \right) \left( \frac{M_P H^{j,\mu \nu 
\rho}}{\mc{V}^{1/3}} \right)
\eea
The canonically normalised Lagrangians are then
\bea
\textrm{Anomalous U(1)s} & & \sqrt{g} H_{\mu \nu \rho} H^{\mu \nu \rho} + \frac{1}{4g^2} F_{\mu \nu} F^{\mu \nu} + \frac{M_P}{\mc{V}^{1/2}} D_2 
\wedge F_2. \\
\textrm{Non-Anomalous U(1)s} & & \sqrt{g} H_{\mu \nu \rho} H^{\mu \nu \rho} + \frac{1}{4g^2} F_{\mu \nu} F^{\mu \nu} + \frac{M_P}{\mc{V}^{2/3}} D_2 
\wedge F_2.
\eea
We conclude that anomalous $U(1)$'s  have a string scale mass. For non-anomalous U(1)s we therefore obtain\footnote{This result agrees with a general 
expression found in \cite{morrison}. For a discussion of volume dependence of
massive $U(1)$s in toroidal compactifications see \cite{antoniadis}.}
\be
M_{U(1)} = \frac{M_P}{\mc{V}^{2/3}}.
\ee
This is the same mass scale of bulk KK modes,
although the mode is associated with local 3-3 strings and has renormalisable couplings to the
matter fields. For intermediate scale models ($M_s \sim 10^{11}$ GeV)  this implies that the non-anomalous $U(1)$'s have a mass of $~10^8$ GeV, 
whereas for TeV
 string scale their mass is of order $~10$ MeV, which should have been observed if their couplings are not too weak

\section{Supersymmetry Breaking in Global Embeddings}
\label{sec4}

We want to study the structure of supersymmetry breaking in this framework. Aspects of this discussion have appeared in
\cite{hepth0610129}, which studied soft terms for local models in the geometric regime
on the assumption of a realistic matter sector.
We will discuss various forms of mediation mechanism and how they can contribute to the soft terms.

\subsection{Gravity Mediation}

As described in e.g. \cite{hepph9707209},
gravity mediation arises from non-renormalisable contact interactions in the supergravity Lagrangian.
The soft terms depend on the F-terms $F^i = e^{\hat{K}/2} K^{i\bar{j}} D_{\bar{j}} W$ and are given by \cite{hepph9707209},
\bea
M_a & = & \frac{F^i \partial_i f_a}{\hbox{Re}(f_a)}, \nonumber \\
m_{\alpha}^2 & = & m_{3/2}^2 - F^i F^{\bar{j}} \partial_i \partial_{\bar{j}} \ln \tilde{K}_{\alpha}, \nonumber \\
A_{\alpha \beta \gamma} & = & F^i \left( \partial_i K - \partial_i \ln (\tilde{K}_{\alpha} \tilde{K}_{\beta} \tilde{K}_{\gamma}) \right).
\eea
The moduli F-terms for the LARGE volume minimum can be directly computed. This computation is described in \cite{hepth0505076, hepth0610129}
and we shall simply state results. Important for this computation is that the matter field kinetic terms behave as
\be
\mc{K} = \frac{C \bar{C}}{\mc{V}^{2/3}} \left(1 + \mc{O}(T_{SM} + \bar{T}_{SM})^{\lambda} + \ldots \right)
\ee

\subsubsection*{Bulk moduli}

The supersymmetry breaking is dominated by the volume modulus. The F-terms have canonical magnitude\footnote{By the
canonical magnitude we mean that
$\vert F^{T_b} \vert = \vert K_{b \bar{b}} F^b \bar{F}^{\bar{b}} \vert^{1/2}$. This differs from
$F^i = e^{K/2} K^{i \bar{j}} D_{\bar{j}} W$ by factors of the K\"ahler metric, which are significant in the large volume limit.
The canonical magnitude is the appropriate quantity to use in dimensional analysis estimates of soft terms, $m_{soft} = F/M_P$.}
$$
\vert F^{T_b} \vert \sim \frac{M_P^2}{\mc{V}}, \qquad \vert F^{T_s} \vert \sim \frac{M_P^2}{\mc{V}^{3/2}},
\qquad \vert F^{S} \vert \sim \frac{M_P^2}{\mc{V}^2}.
$$
The dominant F-term is that of the volume modulus $T_b$, which naively gives non-zero soft terms
of $\mc{O}(M_P/\mc{V}) = \mc{O}(m_{3/2})$. However at leading order the interactions of $T_b$
take the no-scale form,
$$
K = -3 \ln (T_b + \bar{T_b}) + \frac{C \bar{C}}{T_b + \bar{T_b}} + \ldots, \qquad W = W_0 + \ldots.
$$
As is well known and can be easily checked, the induced soft terms vanish for no-scale models.

It is this fact that makes the computation of soft terms in this framework a delicate issue.
The no-scale structure will certainly be lifted at higher order. For example, string loop corrections \cite{hepth0508043}
give rise to corrections to the K\"ahler potential,
$$
K = - 3 \ln (T_b + \bar{T_b}) + \frac{\alpha}{T_b + \bar{T_b}} + \frac{C \bar{C}}{T_b + \bar{T_b}}\left(1 + \frac{\beta}{T_b + \bar{T_b}} \right), 
\qquad W = W_0.
$$
These higher order effects break no-scale and
lead to non-vanishing soft terms. In this case direct evaluation shows the soft masses to be given by
\be
m_Q^2 = m_{3/2}^2 \left( \frac{2(\alpha/3 - \beta) }{(T_b + \bar{T_b})} + \ldots \right).
\ee
Provided there is no cancellation between $\alpha$ and $\beta$, this generates soft masses at order
$$
m_Q^2 \sim \frac{m_{3/2}^2}{(T_b + \bar{T_b})} \sim m_{3/2}^2 \left( \frac{m_{3/2}}{M_P} \right)^{2/3}.
$$
To have  $m_Q\sim$ TeV this would imply a string scale of order $10^{13}$ GeV.
A cancellation between $\alpha$ and $\beta$ is equivalent to the statement that the physical Yukawa couplings do
not depend on $T_b$ even at subleading order in the volume: i.e. that the
expression $\frac{e^{K/2}}{\sqrt{\tilde{K}_{\alpha}\tilde{K}_{\beta}\tilde{K}_{\gamma}}}$ is independent of $T_b$.

While the breaking of no-scale is generic, any soft terms generated in this fashion are
suppressed by factors of volume from the `natural' soft term
scale of $m_{3/2}$. As for $\mc{V} \gg 1$ this represents a very large suppression
it is important to consider all other possible
sources of supersymmetry breaking that may be induced.

The field $T_s$ has $\vert F \vert \sim \frac{M_P^2}{\mc{V}^{3/2}}$ and so would be expected to generate soft terms at the scale
$M_P/\mc{V}^{3/2} = m_{3/2} \left( m_{3/2}/M_P \right)^{1/2}$, whereas $S$ has $\vert F \vert \sim \frac{M_P^2}{\mc{V}^{3/2}}$
and so would give $\vert F \vert \sim \frac{M_P^2}{\mc{V}^{2}}$.
 If we want $F^{T_s}$ to be of order TeV, this would require a string scale of order 
$10^{13}-10^{14}$ GeV. For branes at singularities the dilaton is the gauge
coupling superfield
and so soft terms of this order seem unavoidable. The question is whether there are larger contributions.

In \cite{hepth0505076, hepth0610129} it was assumed that the Standard Model could be realised on a stack of branes wrapping the
cycle $T_s$. As a local model, the coupling of $T_s$ to the Standard Model is then suppressed by only the string
scale (i.e $M_s$ not $M_P$) and
soft terms are generated at order $\vert F^{T_S} \vert/ M_s \sim m_{3/2}$. However, we have seen that due to the clash between chirality and moduli
stabilisation, this is not possible.

\subsubsection*{Local Blow-up Moduli}

Another potential source of soft terms are the local blow-up moduli. These are directly coupled to Standard Model matter.
Being local, they also couple with string scale suppression and if these break supersymmetry they generate direct moduli mediated soft terms.
For explicitness here we focus on the $dP_0$ case where there are two anomalous $U(1)$s.
The twisted sector moduli are charged under the two anomalous U(1)s. These transform as
$$
T_{SM} \to T_{SM} + i Q \lambda, \qquad B_2 \to B_2 + i Q_B \lambda',
$$
where $\lambda, \lambda'$ are the gauge parameters of the two anomalous U(1)s.  There is no loss of generality in making
$T_{SM}$ charged under $U(1)_1$ and $B_2$ charged under $U(1)_2$.
The gauge-invariant K\"ahler potential for these fields is
$$
\mc{K}(T_{SM} + \bar{T}_{SM}) \to \mc{K}(T_{SM} + \bar{T}_{SM} + Q V_{U(1)}),
$$
where $V$ is the vector multiplet for the anomalous $U(1)$. A similar expression holds for $B_2$. On expanding $\int d^4 \theta \mc{K}$, this 
generates both an FI term for the
$U(1)$ and also a mass term for the $U(1)$. The FI term $\xi$ is given by
$$
\int d^4 \theta \, \, \xi V, \qquad \xi = \frac{\partial \mc{K}}{\partial V} \Bigg\vert_{V=0} = Q \frac{\partial \mc{K}}{\partial T} 
\Bigg\vert_{V=0}.
$$
In general supergravity D-terms are given by \cite{hepth0402046}
$$
V_D = \frac{M_P^4}{2 \hbox{Re}(f_a)}  D^a D^a =  \frac{M_P^4}{2 \hbox{Re}(f_a)}  \left( \eta_A^I \partial_I K - 3 r_A \right)^2.
$$
$r_A$ is only non-zero for a gauged R-symmetry and so not relevant.
$\eta_A^I$ is the transformation, $\delta_A \Phi^I = \eta_A^I(\Phi)$.
For moduli charged under an anomalous $U(1)$, $\eta^I = Q$. The D-term for an anomalous $U(1)$ is then
$$
D_{U(1)} = \left( Q \partial_{T_{SM}} K + \sum_i q_i K_i \phi_i  \right),
$$
containing the appropriate Fayet-Iliopoulos term, where $K_i=\partial K/\partial \phi_i$.
We can therefore write the D-term potential for the blow-up moduli as
\be
V = V_{U(1)} + V_{U(1)'} = \frac{1}{Re(S)} \left( Q \partial_{T_{SM}} K + \sum_i q_i K_i  \phi_i  \right)^2
+ \frac{1}{Re(S)} \left( Q' \partial_B K + \sum_i q'_i K_i  \phi_i  \right)^2.
\ee
This admits a minimum at $T = B = \phi = 0$, i.e. at the singular limit.

However, as the superpotential does not depend on $T_{SM}$, this implies
that
\bea
F^{T_{SM}} & = & e^{K/2} K^{T_{SM} \bar{T}_{SM}} D_{T_{SM}} W \nonumber \\
& = & e^{K/2} K^{T_{SM} \bar{T}_{SM}} K_{T_{SM}} W \nonumber \\
& = & 0.
\eea

The essential point is that for the twisted moduli the magnitude of F-term susy breaking equals that of
the FI term. At the singularity itself, the FI term vanishes and thus so does the F-term breaking, since both are proportional to $K_{T_{SM}}$.
 This holds even if we postulate
quantum corrections to the FI term such that the stable supersymmetric configuration is away from the singularity -
the supergravity structure implies that cancellation of the FI term will lead to cancellation of the F-term.

Notice however that in general there are D-flat directions along which the singularity is partially resolved and
the FI term is cancelled against matter field vevs\footnote{In this more generic case the axionic component of $T_{SM}$ mixes with the argument of 
the
complex scalar field that gets a vev, one combination is eaten by the gauge field and the other combination remains
 massless and a candidate to be the QCD axion. This can be seen by
noticing that $(A-\partial a)^2+(A-\partial\theta)^2= 1/2(2A-\partial(a+\theta))^2+1/2(\partial(a-\theta))^2$ which illustrates that for the
 two axions $a$ and $\theta$, the combination $a+\theta $ is eaten by the gauge field whereas the combination $a-\theta$
remains as an effective massless axion. Notice also that the axionic component
 of $B_2$ cannot be a proper axion since it appears explicitly in the DBI action and therefore it does not only have derivative couplings. },
$$
Q \partial_{T_{SM}} K = \sum q_i K_i  \phi_i  \neq 0.
$$
Along this direction $F^{T_{SM}}$ is non-zero and soft terms are generated in the visible sector through the coupling of
$T_{SM}$ to Standard Model matter. Vevving along such a direction is not possible in the $dP_0$ Standard-like models
 but is possible (using the $Z$ direction)
in the $dP_1$ model. In this case, we expect conventional gravity mediation to be realised
through the F-term associated to $F^{T_{SM}}$, except that the soft terms will be suppressed by $M_s$ instead of $M_P$ ($|F^{T_{SM}}|/M_s\sim 
m_{3/2}$).

Thus at the singularity the twisted moduli preserve supersymmetry and do not generate soft terms. If the singularity can be resolved with
non-zero FI term cancelled against the matter field vevs, then soft terms can be generated at $\mc{O}(m_{3/2})$.

\subsection{Anomaly Mediation}

The vanishing of gravity-mediated soft terms may seem to suggest that anomaly-mediation may play an important role.
However the dominant no-scale susy breaking structure implies that anomaly mediated contributions also vanish.
For anomaly mediation in supergravity the gaugino mass formula from \cite{hepth9911029} is
\be
\label{ambp}
m_{1/2} = \frac{g^2}{16 \pi^2} \left[ (3 T_g - T_R) m_{3/2}  + (T_G - T_R) K_i F^i + \frac{2 T_R}{d_R} (\ln \det \tilde{K} \vert_R)_{,i} F^i \right].
\ee
Evaluated for no-scale models, this vanishes. Anomaly mediation is therefore not capable of generating non-vanishing soft terms in this
framework at leading order. Further corrections to $K$ may induce non-vanishing contributions for soft-terms from anomaly mediation, but they will 
have the standard loop factor suppression over the corrections to the standard gravity mediation.

We note that it has recently been argued that the formula (\ref{ambp}) is not fully correct \cite{08010578}, which would lead
to a modification of the soft masses. A detailed discussion of these issues is outside the scope of this paper, and we
refer the interested reader to \cite{08010578} for a fuller discussion of this issue.

\subsection{Gauge Mediation}

A final and possibly surprising source of soft terms is
gauge mediation. To see how this is possible, suppose we have a model such as $dP_1$ where there
exists a non-anomalous U(1) gauge boson that acquire masses through the Green-Schwarz mechanism.
As calculated in section \ref{sec3}, the mass of such a U(1) gauge boson is
$$
M_{U(1)} \sim M_{KK,bulk} \sim \frac{M_P}{\mc{V}^{2/3}}.
$$
The important point is that such U(1) bosons have masses that are less than the local string/KK scale from which RG running should start, and so
represent a threshold in renormalisation group running between the high scale and the weak scale.
Such U(1)s will enter into Feynman diagrams for squark propagators or for the running gauge coupling, and
will affect quantities such as the gauge coupling RGEs.

However, the mass of these $U(1)$s, and thus the threshold scale, is set by the volume, which is the field that
breaks supersymmetry. This implies that (for example) the low-scale gauge coupling depends radiatively on the volume.
At some level therefore such diagrams feel the breaking of supersymmetry and contribute to soft terms in the visible sector.
However such soft terms will vanish both in the limit that gauge couplings go to zero, and also in the limit that the Planck mass is taken
to infinity. They can also only be generated for models with an appropriate non-anomalous $U(1)$.

\subsection{Summary}

Let us summarise this section on supersymmetry breaking. At the singularity,
all leading order contributions to the soft terms (both gravity and anomaly-mediation) vanish.
If some of the twisted moduli are vevved against matter fields, then these can break supersymmetry and generate non-vanishing
soft terms of $\mc{O}(m_{3/2})$. Otherwise, the leading contributions to soft terms seems to come from sub-leading terms
that are difficult to calculate and are volume-suppressed compared to $m_{3/2}$.

In all cases, the `mirror mediation' mechanism for flavour universality is at work. The complex structure moduli, responsible for flavour play a 
subdominant role in supersymmetry breaking. Furthermore in the simplest del Pezzo singularities, the structure of the Yukawas is actually rigid and 
not even the complex structure
moduli appear. Therefore approximate flavour universality is guaranteed in this scenario.

\section{Conclusions}

We have made progress into closing  the gap between the local phenomenological
D-brane models and the important issue of moduli stabilisation and supersymmetry breaking.
Our main results can be summarised as follows.

\begin{itemize}

\item{} We generalised the previous constructions of realistic models on branes at singularities in the bottom-up approach to model building
 using recent developments in terms of quiver theories.
The constructions can be generalised beyond orbifold singularities to include higher order del Pezzo  and, in principle, infinite classes
of singularities
such as $Y_{p,q}$ and $L_{l,m,n}$.
We used  triplication of families and especially the global symmetries of the Yukawa couplings to select preferred models.
We found that the main problem of $dP_0$ models, namely that the D3-D3 couplings lead to a mass matrix with eigenvalue $(M,M,0)$ clearly against 
observations, is naturally solved if we move to more general singularities such as $dP_1$ for which the eigenvalues are of the form $(M,m,0)$ with 
$m$ hierarchically smaller than $M$, while keeping the other attractive properties of the
models such as triplication of families.

\item{} We argued that the LARGE volume scenario of moduli stabilisation {\it implies} the bottom-up approach of model building.
This is relevant because the original justification of this approach
was on the basis of simplicity of model construction. Now we can see that this is required from a concrete mechanism of moduli stabilisation.

\item{}
Being in the vicinity of a singularity implies a particular form of the
local effective field theory. In particular Yukawa couplings do not vanish
(contrary to the blown-up case).
Also, the volume dependence of
K\"ahler potentials for matter fields and blown-up modes were determined
as well as an explicit derivation of the mass of  anomalous and non-anomalous $U(1)$s.
The generic existence of hyperweak interactions coming from D7 branes wrapping the large cycle was emphasised and their potential
phenomenological implications were
outlined, including their potential relevance to recent CDF results. A more detailed study is left for the future.

\item{} We outlined the minimal requirements to extend our model to a
local  compact Calabi-Yau compactification with moduli stabilised. In particular avoiding global/local
 mixing requires the Standard Model modulus to be fixed by a combination of D-terms and loop corrections to
 K\"ahler potential rather than the standard non-perturbative effects. This leaves the axionic partner of the blow-up mode
(or a combination of this and the argument of the complex scalar field that gets a vev to cancel the FI contribution to the D-terms)
 unfixed which could play the role of QCD axion. An explicit construction for a compact model with all tadpoles cancelled is left for future work.

\item{}
The structure of soft supersymmetry breaking terms and their scales were estimated.
 In particular they are different from previous studies and
 require further investigation. Even though, as argued in the
 introduction, supersymmetry breaking requires a full global analysis,
 we have seen that due to the bulk no-scale structure the leading order contribution for soft
 supersymmetry breaking terms may be local. The F-term of the modulus
 corresponding to the Standard Model cycle may be responsible for the soft terms and its
 mediation is only supressed by $M_s$ instead of $M_P$. Other
 possible sources of soft terms are also allowed depending on the
 value of the blow-up mode at the minimum of its potential. However a full
 phenomenological analysis of soft supersymmetry breaking is beyond
 the scope of this article.

\item{} Even though our models were restricted to branes at singularities,
many of our results should extend to more general compactificatons
such as F-theory constructions of GUT models  \cite{08023391,vafaex} which are also local. It is an interesting open question to make the connection 
directly between the LARGE volume scenario of moduli stabilisation and    F-theory models, although
direct connection with concrete local IIB orientifold models in terms of magnetised D7 branes should be tractable.

\end{itemize}

\appendix

\section{Non-anomalous U(1) in the $dP_{1}$ quiver}

We summarize
the $U(1)$ charge assignments of fields in the $dP_{1}$ quiver in the table below. We denote the 73 strings streching between the D7 node $m_{i}$ and
D3 node $n_{j}$ as $A^{ij}$.


         With these charge assignments it can be checked that the
combination  $Q = \sum { Q_{ni} \over n_{i} }$ is indeed anomaly free,
i.e the condition for absence of gravitational, $U^{3}(1)$ and mixed anomalies
(involving all the non-abelian gauge groups) are satisfied
\begin{equation}
    \sum_{\rm species} Q = 0
\end{equation}
\begin{equation}
  \sum_{\rm species} Q^{3} = 0
\end{equation}
\begin{equation}
  \sum_{\rm rep} Q {\rm{Tr}} (T^{a} T^{b}) = 0
\end{equation}

\begin{center}
\begin{tabular}
{ c c c c c c}
 Field  & $Q_{n1}$  & $Q_{n2}$ & $Q_{n3}$ & $Q_{n4}$ & $Q = \sum { Q_{ni} \over n_{i} }$\\
\hline
$\Phi$ & 0 & 0 & +1 & -1 & $+{1 \over n_{3} } -{1 \over n_{4} } $\\
$Z_{K}$ &-1 & 0 & 0 & +1 &  $- {1 \over n_{1} } + {1 \over n_{4} } $ \\
$X_{i}$ & +1 & -1 & 0 & 0 & $+{1 \over n_{1} } - {1 \over n_{2} } $ \\
$X_{3}$ & +1 & -1 & 0 & 0 & $+{1 \over n_{1} } - {1 \over n_{2} } $ \\
$Y_{j}$ & 0 & +1 & -1 & 0 & $+{1 \over n_{2} } - {1 \over n_{3} } $  \\
$Z_{3}$ & -1 & 0 & +1& 0 & $-{1 \over n_{1} } + {1 \over n_{3} } $ \\
$Y_{3}$ & 0& +1 & 0 & -1 &  $+{1 \over n_{2} } - {1 \over n_{4} } $\\
$A^{11}$ & -1 & 0 & 0 & 0 & $-{1 \over n_{1}}$  \\
$A^{12}$ & 0 & +1 & 0 & 0 & $+{1 \over n_{2}}$  \\
$A^{22}$ & 0 & -1 & 0 & 0 &  $-{1 \over n_{2}}$\\
$A^{23}$ & 0 & 0 & +1 & 0 &   $+{1 \over n_{3}}$\\
$A^{33}$ & 0 & 0 & -1 & 0 &   $-{1 \over n_{3}}$\\
$A^{34}$ & 0 & 0 & 0 & +1 &  $+{1 \over n_{4}}$ \\
$A^{44}$ &0 & 0 & 0 &  -1 & $-{1 \over n_{4}}$ \\
$A^{41}$ & +1 & 0 & 0 & 0 &  $+{1 \over n_{1}}$ \\
$A^{53}$ & 0 & 0 & -1 & 0 &  $-{1 \over n_{3}}$ \\
$A^{51}$ & 1 & 0 & 0 & 0 &  $+{1 \over n_{1}}$ \\
$A^{62}$ & 0 & -1 & 0 & 0 & $-{1 \over n_{2}}$  \\
$A^{64}$ & 0 & 0 & 0 & +1 & $+{1 \over n_{4}}$  \\
\hline
\end{tabular}
\end{center}

\acknowledgments{}
We are indebted to Angel Uranga for many patient explanations about singularities and many useful discussions on the subject of this work. We
are also grateful for
conversations with S. AbdusSalam,
B. Allanach, M. Cicoli, M. Kreuzer, F. Marchesano, K. Narayan, E. Palti,
H. Reall, S. Theisen and M. Schulz.
JC thanks the Royal Society for a University Research Fellowship.
FQ is partially funded by STFC and a Royal Society Wolfson award.
 AM and FQ thank the CERN theory group for hospitality during the `String Phenomenology Institute'. FQ also thank the organisers and participants of 
the `Stringy Reflections on the
 LHC' at the Clay Mathematics Institute,
for hospitality and many useful discussions.

\end{document}